\documentclass[oldversion]{aa}
\usepackage{graphicx}
\usepackage{txfonts}
\usepackage{lscape}
\usepackage{multirow}
\usepackage[authoryear]{natbib}
\begin{document}

    \renewcommand{\topfraction}{0.85}
    \renewcommand{\bottomfraction}{0.7}
    \renewcommand{\textfraction}{0.15}
    \renewcommand{\floatpagefraction}{0.66}

\title{A search for VHE counterparts of Galactic \emph{Fermi} bright sources and MeV to TeV spectral characterization}

\small{
\author{P.H.T.~Tam \inst{1} \thanks{now at Institute of Astronomy and Department of Physics, National Tsing Hua University, Hsinchu 30013, Taiwan}
 \and S.J.~Wagner \inst{1}
 \and O.~Tibolla \inst{2}
 \and R.C.G.~Chaves \inst{2}
}

\institute{
Landessternwarte, Universit\"at Heidelberg, K\"onigstuhl, D 69117 Heidelberg, Germany \\
\email{ phtam@lsw.uni-heidelberg.de }
\and
Max-Planck-Institut f\"ur Kernphysik, P.O. Box 103980, D 69029
Heidelberg, Germany
}

\abstract
{Very high-energy (VHE; E$>$100~GeV) $\gamma$-rays have been detected from a wide range of
astronomical objects, such as pulsar wind nebulae (PWNe), supernova remnants (SNRs), giant molecular clouds, $\gamma$-ray binaries, the Galactic Center, active galactic nuclei (AGN), radio galaxies, starburst galaxies, and possibly star-forming regions as well. At lower energies, observations using the Large Area Telescope (LAT) onboard \emph{Fermi} provide a rich set of data which can be used to study the behavior of cosmic accelerators
in the MeV to TeV energy bands. In particular, the improved angular resolution of current telescopes in both bands compared to previous instruments significantly reduces source confusion and facilitates the identification of associated counterparts at lower energies. In this paper, a comprehensive search for VHE $\gamma$-ray sources which are spatially coincident with Galactic \emph{Fermi}/LAT bright sources is performed, and the available MeV to TeV spectra of coincident sources are compared. It is found that bright LAT GeV sources are correlated with TeV sources, in contrast to previous studies using EGRET data. Moreover, a single spectral component seems unable to describe the MeV to TeV spectra of many coincident GeV/TeV sources. It has been suggested that $\gamma$-ray pulsars may be accompanied by VHE $\gamma$-ray emitting nebulae, a hypothesis that can be tested with VHE observations of these pulsars.}
   \maketitle

\section{Introduction}

Our understanding of the very high-energy (VHE; E $>$100 GeV) sky has greatly improved during the last few years, thanks to the high sensitivity of current imaging atmospheric Cherenkov telescopes (IACTs), e.g. H.E.S.S., MAGIC, and VERITAS. They typically cover the energy range of $\sim$100~GeV up to several tens of TeV, and provide an
angular resolution of $\sim 6\arcmin$. This allows spectral and morphological studies of the various types of VHE sources: pulsar wind nebulae (PWNe), supernova remnants (SNRs), giant molecular clouds, $\gamma$-ray binaries, the Galactic Center, active galactic nuclei (AGN), radio galaxies, starburst galaxies, and possibly star-forming regions as well. See~\citet{VHE_Review_08} for a review of the field in 2008, with more recent updates given
by the H.E.S.S. \citep{hess_survey09}, MAGIC \citep{MAGIC_results_Fermi_Sym09}, and VERITAS collaborations \citep{VERITAS_results_Fermi_Sym09,VERITAS_cygnus_survey09}. However, many of the sources are not yet identified at other wavelengths, e.g., nearly a third of the Galactic H.E.S.S. sources have no firm identification; in many cases, there are multiple plausible counterparts while, in other cases, no viable counterparts have yet been identified.

Gamma-ray observations of Galactic sources can help us to solve a number of important astrophysical questions, including (1) the physics of pulsars, PWN and SNR; and (2) the origin of cosmic rays. Our Galaxy contains a large number of cosmic accelerators, where particles are accelerated to highly-relativistic energies (up to at least $10^{14}$~eV). The origin of cosmic rays is still not well known, largely because of the lack of directional information of these particles. These very energetic particles can be traced within our Galaxy by a combination of non-thermal X-ray emission and $\gamma$-ray emission via leptonic (such as inverse Compton scattering of electrons, Bremsstrahlung and synchrotron radiation) or hadronic (via the decay of charged and neutral pions, due to interactions of energetic hadrons) processes. Therefore, observations of $\gamma$-rays at energies $\ga$100~MeV can probe the sources of particle acceleration.

The Large Area Telescope (LAT), onboard the \emph{Fermi Gamma-ray Space Telescope}, provides the best information of the non-thermal sky in the energy range from 20 MeV to 300 GeV. The point-source sensitivity of LAT is $\sim$10$^{-8}\,\mathrm{ph}\,\mathrm{cm}^{-2}\,\mathrm{s}^{-1}$ above 100~MeV in one year of survey-mode observations~\citep{lat_technical}, which is an order of magnitude better than that of its predecessor, the Energetic Gamma Ray Experiment Telescope (EGRET). Its angular resolution is $\la 0\fdg6$ above 1~GeV, which is particularly important for identifying $\gamma$-ray sources with multi-wavelength counterparts and revealing their nature~\citep{lat_technical}. As an important step towards the first source catalog, the LAT collaboration has published a bright source list (BSL) that includes 205 sources, designated with the prefix 0FGL, using data taken during the first three months of observations~\citep{bsl_lat}. Among them, 121 sources are identified with AGN and one with the Large Magellanic Cloud. Most of the remaining 83 sources are believed to originate in our own Galaxy. It is natural to investigate which of them also have been detected at energies $\ga$100~GeV.

The search for VHE counterparts of LAT sources is important for the following reasons:
\begin{enumerate}
  \item it aids the identification of the true nature of the LAT sources through their VHE counterparts;
  \item for pulsars, it helps us to identify their VHE-emitting nebulae;
  \item it may provide us with broad-band $\gamma$-ray spectra, thereby better constraining the emission mechanisms (e.g. distinguish between hadronic and leptonic scenarios).
\end{enumerate}

\citet{funk08} compare $\gamma$-ray sources in the third EGRET (3EG) catalog~\citep{3rd_egret_cat} and the 22 H.E.S.S. sources known at the time within the region of $l=-$30$\degr$ to 30$\degr$, $b=-$3$\degr$ to 3$\degr$~\citep{hess_survey}. They do not find spatial correlation between the two populations. Though some coincidence cases are found, the authors conclude that these few cases can be explained by chance coincidence. However, due to the capabilities of EGRET, this study suffers from the following limitations: (1) The sensitivity of EGRET is lower than that of LAT. The lack of photon statistics leads to poorly constrained spectral indices and the spectra terminate $\la$10~GeV at the upper end for a typical source; (2) EGRET sources are only localized at degree scales, which is much larger than the angular resolution of IACTs. The second point is the instrumental reason which explains the weak correlation of EGRET and H.E.S.S. sources~\citep{funk08}. These
shortcomings are now largely overcome due to the enhanced performance of LAT over EGRET. In addition to the above caveats, they do not consider the extension of the VHE $\gamma$-ray sources in their analysis. As such, the full potential of this search is not realized for very extended sources like the SNR RX~J1713.7$-$3946, as pointed out by \citet{HowCanFermiHelp_SciNeGHe08}. After the launch of LAT, one largely benefits from the increased LAT angular resolution over previous studies. As noted in~\citet{lat_technical}, EGRET could not distinguish the GeV emission of RX~J1713.7$-$3946 from 3EG~J1714$-$3857, while the capabilities of LAT allow one to study individual sources in this region, which contains three VHE $\gamma$-ray sources~\citep[See Fig. 1 in][]{CTB_37A}.

The water Cherenkov detector MILAGRO covers the energies above $\sim$1~TeV and its angular resolution can reach $<$ 1$\degr$. Using MILAGRO, a search for $\gamma$-rays from the Galactic LAT BSL has been performed by~\citet{milagro_bsl}. They found that 14 sources (of the selected 34) show evidence for multi-TeV $\gamma$-ray emission at a significance of $\geq3\sigma$, although most of the source candidates cannot be established as firm detection on an individual basis~\citep{milagro_bsl}.

In this paper, a search for VHE counterparts of all the presumed Galactic sources in~\citet{bsl_lat} and \citet{lat_psr_cat} is performed, with spatial coincidence as the primary criterium for association. The extensions of the VHE $\gamma$-ray sources are taken into account and the search is not limited to the H.E.S.S. Galactic plane survey region. The broad-band MeV to TeV spectra of coincident sources are then presented.

\section{Search of spatial coincidence}

\subsection{The \emph{Fermi} and VHE catalogs}

\citet{bsl_lat} present 205 point-like sources which were detected at or above the 10-$\sigma$ level in the 0.2--100 GeV band, based on three months of observations (August~4, 2008 -- October~30, 2008). The list is not flux-limited and thus is not uniform. The following information is given for each source: its position, positional uncertainty (95\% confidence level, C.L.), significance, flux in two energy bands (100~MeV--1~GeV and 1--100~GeV), and whether or not there is evidence for variability over the above-mentioned period. In addition, \citet{bsl_lat} assign for each source its source class, as well as $\gamma$-ray and lower energy association (if any). Those sources that are classified as extragalactic (all AGN and the Large Magellanic Cloud) are not considered in this work.

The remaining source list contains 83 sources, comprising 15 radio/X-ray pulsars, 15 pulsars newly discovered by the LAT, two high-mass X-ray binaries (HMXBs), one globular cluster (47~Tucanae), 13 SNR/PWN candidates\footnote{possibly associated with SNR or PWN, but the emission may come from unknown pulsars}, and 37 sources without obvious counterparts at lower energies \citep[among them the Galactic Center;][]{bsl_lat}. \citet{lat_psr_cat} presents the first LAT pulsar catalog. Those 16 pulsars that are not present in the above BSL are also included in this study. Therefore, most of the LAT bright sources considered in this work should be Galactic in origin.

The number of VHE $\gamma$-ray sources is larger than 50 as of Fall 2009~\citep{VHE_Review_08,hess_survey09,MAGIC_results_Fermi_Sym09,VERITAS_results_Fermi_Sym09,VERITAS_cygnus_survey09}. Galactic sources summarized in the above publications are used in the coincidence search in this work. Therefore,
our comparison is based on published sources only. The VHE $\gamma$-ray source positions and source extensions in this work are taken from the corresponding publications shown in Tables~\ref{tab:pos_coincidence1}, \ref{tab:pos_coincidence2}, and \ref{tab:pos_coincidence3}. At higher energies, the MILAGRO collaboration reported evidence for multi-TeV $\gamma$-ray emission from several LAT source positions~\citep{milagro_bsl}. Only those source candidates with a significance larger than 5$\sigma$ are regarded as TeV sources here and are included in this study\footnote{For example, HESS~J1833-105~\citep{icrc07_1833_105}, having a significance of 4.0$\sigma$ only but included in \citep{lat_psr_cat} as a counterpart of LAT pulsar PSR~J1833-1034, is not considered here.}. With several tens of known sources in both the GeV and TeV bands, a systematic cross-correlation study is conducted.

\subsection{Level of spatial coincidence}
\label{position_procedure}

To quantify the level of spatial coincidence, the following scheme is employed. Let $d$ be the distance between a centroid position best fit by LAT and the best-fit centroid of a nearby VHE $\gamma$-ray source. The radius of the 95\% confidence region for the LAT source is the uncertainty on the centroid position of the given LAT source, which is typically $\sim0\fdg1$. Most VHE $\gamma$-ray sources are extended, with a typical size of $0\fdg1-0\fdg5$. Let $e$ be the sum of the radius of the 95\% confidence region and the size of the VHE $\gamma$-ray source.

For each LAT source, if a VHE $\gamma$-ray source was found where $d - e < 0$, the source pair is classified as a spatially coincident case (i.e. category \emph{Y} -- Yes). Given the possible extended nature of many LAT bright sources, a category \emph{P} (for Possible) is defined for pairs where $0 < d - e < 0.3\degr$, so that the sources seen by LAT and the VHE instruments may actually overlap each other; they are possibly coincident cases. If no reported VHE $\gamma$-ray source was found with $d - e < 0.3\degr$, the LAT source falls into the coincidence level \emph{N} (for No), i.e., no coincidence with any VHE $\gamma$-ray source. If there are several nearby VHE $\gamma$-ray sources, only the closest VHE $\gamma$-ray source would be considered.

\subsection{Spatial coincidence GeV/TeV pairs}
\label{position_results}

In the search, 24 coincident cases (\emph{Y}, among them two are coincident with MILAGRO source only) and 7 possibly-coincident cases ({\it P}) are found. The results are presented in Tables~\ref{tab:pos_coincidence1}, \ref{tab:pos_coincidence2}, and \ref{tab:pos_coincidence3}. No reported VHE $\gamma$-ray sources are found in the remaining 68 sources.

According to the nature of the LAT sources, the results are summarized as follows:
\begin{enumerate}
  \item Eight LAT pulsars are spatially coincident with a source detected using IACTs, which may be the VHE-emitting PWN. There are two additional pulsars which are possibly coincident with an IACT source. Two others have a MILAGRO counterpart, but have not yet been detected by IACTs.
  \item Among the 13 SNR/PWN candidates in the \emph{Fermi} BSL, more than a half (7) are spatially coincident with a VHE $\gamma$-ray source, and another one is a possibly coincident case. The seemingly high fraction of coincidence is partly due to a better coverage of the inner Galaxy region, where most SNR/PWN candidates are located. This results in a generally better sensitivity for this class of sources than for other classes.
  \item The two HMXBs listed in the BSL (0FGL~J0240.3$+$6113/LS~I~$+$61~303 and 0FGL~J1826.3$-$1451/LS~5039) are both found to be spatially coincident with a VHE $\gamma$-ray source, identified with the same corresponding binary.
  \item Five of the 37 unidentified 0FGL~sources are spatially coincident with a VHE $\gamma$-ray source. The number increases to nine if possibly coincident cases are included.
\end{enumerate}

In addition, we note that a new VHE source near PSR~J1119$-$6127 has been announced in an oral presentation\footnote{See http://cxc.harvard.edu/cdo/snr09/pres/DjannatiAtai\_Arache\_v2.pdf}, but it has not been published with any written reference to our knowledge. Given that the best-fit centroid and extension were not given, we do not treat it as an entry in our sample\footnote{In the first LAT catalog, that can be found at http://fermi.gsfc.nasa.gov/ssc/data/access/lat/1yr\_catalog/, the authors claim that there is an association of the LAT source 1FGL~J1119.4-06127 with a VHE source, designated HESS~J1119$-$614, which may be the same VHE source.}.

With such a large number of coincident cases, the relationship between the GeV and TeV sources is explored. In the next section, the $\gamma$-ray spectral energy distributions (SEDs) are constructed for those coincident and possibly coincident GeV/TeV source pairs with published VHE spectrum.

\section{The $\gamma$-ray spectral energy distributions}

\subsection{Construction of PL spectrum in the LAT energy range}
\label{Sect:SED_construction}

\citet{bsl_lat} provide photon fluxes and respective errors in two energy bands: low energy (100~MeV--1~GeV) and high energy ($1-100$~GeV). Since photon spectra are not given in the BSL, we attempt to estimate the spectra of individual sources.

Assuming that a single pure power law (PL) represents the spectrum in the LAT energy range, the photon flux in the low (10$^2$--10$^3$~MeV) and high energy  (10$^3$--10$^5$~MeV) bands, respectively, are given by $F_{23}  = k \int_{0.1}^1 E^{-\Gamma} dE$ and $F_{35}  = k \int_1^{100} E^{-\Gamma} dE$, where $E$ is measured in GeV, $\Gamma$ is the photon index, and $k$ is the normalization at 1~GeV. From these two expressions, $k$ and $\Gamma$ can be calculated. Using the available flux errors ($\Delta F_{23}$ and $\Delta F_{35}$), uncertainties in $k$ and $\Gamma$ ($\Delta k$ and $\Delta\Gamma$) are obtained by error propagation. The spectra are then constructed in the form of ``bowties''. For those where $F_{23}$ is given as a 2-$\sigma$ upper limit, the calculated $\Gamma$ can be treated as an upper limit and the reconstructed spectra can be seen as the ``softest possible'' PL spectra. The PL spectra are plotted from 100~MeV up to a certain maximum energy, $E_\mathrm{max}$ ($\leq 100$~GeV), which is defined by requiring that the photon spectrum above $E_\mathrm{max}$ contains 10 photons over the three months of observations\footnote{using the LAT on-axis effective area above 1~GeV of $\sim$8000~cm$^2$ and a mean on-axis exposure of $\sim$1~Ms~\citep{bsl_lat}}. This results in a range of values for $E_\mathrm{max}$ from $\sim$3~GeV to 100~GeV. The single PL assumption does not hold in general. Given the limited information available in the BSL, such an assumption should be seen as a very rough estimation of the source spectra and it is used in this work for a visual GeV/TeV spectral comparison. A cut-off between the GeV and TeV bands has been measured for several sources including pulsars. Therefore, we also plot the best-fit spectra when a detailed LAT spectrum is available in the literature (Vela, Crab, Geminga, PSR~J1706$-$44, and LS~I~$+$61~303). For the cases of 0FGL~J0617.4$+$2234 and 0FGL~J1746.0$-$2900, the double PL spectra derived for 3EG~J0617$+$2238 and 3EG~J1746$-$2851, respectively, by~\citet{Bertsch00} are also shown for comparison.

\subsection{The MeV--TeV SEDs}
\label{Sect:SED_results}

The sources considered here are those 0FGL/VHE pairs with spatial coincidence levels {\it Y} and {\it P} and with VHE spectral information available in the literature. In the case of HESS~J1923$+$141 where only a VHE flux is given, a typical spectral index is assumed. In addition, there are two pulsars for which a MILAGRO candidate counterpart is reported but there is no VHE $\gamma$-ray detection using IACTs (see Table~\ref{tab:pos_coincidence2}).

SEDs of the 28 cases in the energy range from 100~MeV to $>$1~TeV are depicted in Figs.~\ref{EGR_psr} to \ref{GC}. Systematic errors in spectral indices and normalization are not shown, which for TeV spectra are $\sim$20\% for most sources and for GeV spectra are $\sim$20--30\% \citep[the latter is inferred from flux estimation systematics in][]{bsl_lat}. An overall inspection of the SEDs immediately shows that single PLs from 200~MeV to $\sim$10~TeV cannot describe most GeV--TeV $\gamma$-ray spectra. This is not surprising given the large range in photon energy (i.e. five orders of magnitude), as no photon spectrum from any emission mechanism is expected to be unbroken for such a large energy span. The only example for which a pure PL may still work is 0FGL~J1836.1$-$0727/HESS~J1837$-$069, which is a possibly coincident pair (\emph{P}). The most common board-band behaviors are a cut-off at energies below $\sim$100~GeV (dominating in the pulsar class), and a spectral break between the LAT and the VHE bands (dominating in the unidentified LAT sources).

The SEDs of the LAT source classes including pulsars, SNR/PWN candidates, and unidentified $\gamma$-ray sources are presented in this section. LS~I~$+$61~303 and the Galactic Center region are discussed in Sect.~\ref{Sect:LSI} and \ref{Sect:GC}, respectively.

\subsubsection{Pulsars}

Figure~\ref{EGR_psr} shows the four $\gamma$-ray pulsars known in the EGRET era, Fig.~\ref{radio_psr} shows the four radio pulsars first detected in $\gamma$-rays by LAT, and Fig.~\ref{LAT_psr} shows the three new pulsars after a blind search for pulsations in the LAT data~\citep{latpsr_blind}. Besides the Crab, no off-pulse emission is found in the LAT data of the other 10 pulsars, suggesting that most of the emission from pulsars seen with LAT is pulsed and originates from the pulsars themselves. On the other hand, extended regions are seen at energies above 100~GeV in these 10 cases (except for the Crab, which appears point-like to all IACTs). Their VHE emission ($>$100~GeV) is unpulsed and for many of them (e.g. Vela X) this emission have been attributed to PWNe, although in some cases there exists other possibilities to explain the VHE $\gamma$-ray source (e.g. a spatially coincident SNR).

The SEDs of the pulsars essentially depict the pulsed component in the LAT energy band and the unpulsed component in the VHE band. It is suggested that the emission below and above $\sim$100~GeV comes from two different emission regions, e.g. pulsed emission from the pulsar magnetosphere and unpulsed emission from the PWN. It can be seen that (1) a cut-off must exist between the LAT ``bowties'' and the corresponding VHE spectra (with the notable exception of the Crab) -- this is demonstrated with a detailed spectral study of pulsars \citep[e.g.,][]{lat_psr_cat}; (2) the energy output at GeV energies is at least an order of magnitude higher than that in the VHE band. This indicates that for the pulsar population presented in this section, the PWNe radiate less energy than the $\gamma$-ray pulse emitting regions.

However, the PL derived LAT spectra are not always good representations of the reported spectra for individual sources. This is demonstrated in Fig.~\ref{EGR_psr} where both the ``bowtie'' spectra and the derived spectra in~\citet{lat_psr_cat} are shown. In all the other cases, only the spectra as presented in~\citet{lat_psr_cat} are depicted.

\subsubsection{SNR/PWN candidates}

The SEDs of those 0FGL sources classified as SNR/PWN candidates are shown in Figs.~\ref{SNR1} and \ref{SNR2}. The GeV--TeV spectral connection varies among the sources in this class. The TeV spectra are not simply the PL tails of the GeV spectra. There are cases where the extrapolation of the LAT ``bowtie'' to TeV energies is at least an order of magnitude higher than the measured VHE flux level (e.g. the spatially coincident case 0FGL~J1801.6$-$2327/HESS~J1801$-$233, a cut-off occurs between the two energy bands), while for another coincident case (0FGL~J1834.4$-$0841/HESS~J1834$-$087) the PL extrapolation to the VHE band is below the measured VHE level and a second spectral component above $\sim$200~GeV is needed to explain the TeV excess.

There is only one case (0FGL~J0617.4$+$2234) where a broken PL describes the LAT spectrum better than a single power law. The ``bowties'', which are derived \emph{a priori} from PLs, may be closer to the real spectra compared to the case of pulsars. If that is the case for 0FGL~J1801.6$-$2327/HESS~J1801$-$233 and 0FGL~J1923.0$+$1411/HESS~J1923$+$141, a spectral break may occur at energies in the largely unexplored energy range of 10--100~GeV for these two sources\footnote{The LAT spectrum for 0FGL~J1801.6$-$2327 is the softest possible PL, while the HESS~J1923$+$141 spectrum is derived assuming a PL index $\Gamma=2.8$.}.

\subsubsection{Unidentified LAT sources}

The SEDs of those 0FGL~sources without obvious counterparts are shown in Figs.~\ref{Unid1} and \ref{Unid2}. There is so far no published spectra of this LAT source class. For the case of 0FGL~J1839.0$-$0549/HESS~J1841$-$055, the spectrum may span from $\sim$100~MeV to ~$\sim$80~TeV, with a possible break within or close to the ``energy gap'' at $\sim$60--500~GeV. If the GeV and TeV sources are indeed associated, they might represent a group of ``dark accelerators'' which have a broad $\gamma$-ray spectrum. All SEDs are consistent with the assumption that a spectral break exists between the two energy bands, except for the case of 0FGL~J1805.3$-$2138/HESS~J1804$-$216, a spatially coincident case ({\it Y}).


\subsection{Comparison of the flux and photon indices in the GeV and TeV energy bands}

A comparison of the flux levels in the GeV and TeV energy bands for coincident GeV/TeV sources (category \emph{Y}) is depicted. Figure~\ref{F23} shows the photon flux in the 100~MeV -- 1~GeV band plotted against that in the 1--10~TeV band (derived according to Sect. \label{Sect:SED_construction}). For most sources, the photon flux in the 1--10~TeV band, $F_{1-10\,\mathrm{TeV}}$, is about $10^{-5}$ to $10^{-6}$ that in the 0.1--1~GeV band. Figure~\ref{F35} shows the photon flux in the 1--100~GeV band plotted against that in the 1--10~TeV band. For most sources, photon flux in the 1--10~TeV band, $F_{1-10\,\mathrm{TeV}}$, is about $10^{-4}$ to $10^{-5}$ that in the 1--100~GeV band.

Figure~\ref{Gamma} depicts the photon indices in the 0.1--100~GeV band derived according to Sect.~\ref{Sect:SED_construction} against photon index in the 1--10~TeV band. It can be seen that the TeV spectra are similar to or harder than the GeV spectra for most sources, i.e. $0\lesssim(\Gamma_{1-10\,\mathrm{TeV}}-\Gamma_{0.1-100\,\mathrm{GeV}})\lesssim1$.

\section{Notes on selected sources}
\label{Sect:individual_srcs}

Although detailed analysis of the LAT data for each individual source is beyond the scope of this paper, some comments on the following sources are given.

\subsection{Crab pulsar and nebula}

The Crab pulsar and nebula are among the best-studied non-thermal objects in the sky. The pulsed emission above 100~MeV and up to $\sim$10~GeV is clearly detected with LAT. Two strong peaks are seen in the phase histogram. A spectral fit of the pulsed emission using a PL with an exponential cut-off gives a cut-off energy of $\sim$8.8~GeV~\citep{lat_crab}. There is evidence of pulsed emission up to $\sim$25~GeV, as measured using the MAGIC telescope~\citep[see Fig.~\ref{EGR_psr},][]{magic_crab_psr}. The flux reported by MAGIC is consistent with the exponential cut-off in the spectrum measured by LAT.

Evidence for unpulsed emission was already present in the EGRET data~\citep{egret_crab_nebula}. The LAT measurement of this component can be well fit by a single PL with $\Gamma\sim1.9$ up to $\sim$300~GeV. This unpulsed spectrum agrees well with the VHE spectra measured by the IACTs MAGIC, H.E.S.S. and VERITAS~\citep{lat_crab}. In particular, there appears to be a deviation from a pure PL in the MAGIC spectrum below $\sim$100~GeV~\citep{magic_crab_nebula}.

\subsection{Vela pulsar and Vela X}

The Vela pulsar is the strongest persistent GeV source and was the first target of LAT observations. The complex pulse profile is dominated by two peaks with a pronounced ``bridge'' between them. The phase-averaged spectrum, which is essentially the pulsed emission, can be well described by a PL with an exponential cut-off at $\sim$2.9~GeV. The off-pulse emission is much weaker, and a 95\% C.L. upper limit of the photon flux of $1.8\times10^{-7}$cm$^{-2}$~s$^{-1}$ is derived at the pulsar position in the 0.1--10~GeV band \citep[shown in Fig.~\ref{EGR_psr},][]{lat_velapsr}.

To the south of the pulsar, an extended VHE $\gamma$-ray source spatially coincident with the Vela~X region, HESS~J0835$-$455, has been detected. The observations represent the first measurement of a SED peak in a VHE $\gamma$-ray source~\citep{hess_velaX}. The PL with exponential cut-off fit of this PWN is reproduced in Fig.~\ref{EGR_psr}. An analysis of the Vela~X region does not establish a nebula component based on the first three months of LAT observations~\citep{lat_velaX}.

\subsection{Geminga}

The Geminga pulsar is the first known radio-quiet $\gamma$-ray pulsar in the sky~\citep{Geminga_psr_discovery}. See Fig.~\ref{EGR_psr} for its SED. While EGRET data are well fit by a single PL up to 2~GeV \citep[but shows evidence of a cut-off above 2~GeV;][]{egret_geminga}, the cut-off energy is determined to be $\sim$2.6~GeV using the first seven months of LAT data~\citep{lat_geminga}. There appears to be an excess at $\sim$20~GeV when compared to the fit with a PL with exponential cut-off. The reason may be due to low statistics or effects of the fitting method, but it might also indicate a seperate and harder spectral component~\citep{lat_geminga}. There is as yet no evidence for unpulsed emission.

Evidence for multi-TeV emission around the pulsar was reported in the MILAGRO survey of the Galactic plane~\citep{milagro_GPS} and in the search for MILAGRO counterparts of \emph{Fermi} sources~\citep{milagro_bsl}, using a point source analysis, at $\sim$3$\sigma$ (post-trial) significance levels. Assuming that the emission is extended, the significance increases to 6.3$\sigma$ at the position of the pulsar. If the detection is real, the size of the MILAGRO emission is $\sim$2$\fdg$6. At a distance of only $\sim$250 pc, this extent is similar to more distant PWN~\citep{milagro_bsl}. On the other hand, VERITAS observations resulted in no detection but rather a 99\% C.L. flux upper limit (above 300~GeV) of $2\times10^{-12}$~cm$^{-2}$~s$^{-1}$, assuming point source emission from the pulsar~\citep{veritas_geminga}. Although IACTs suffer from reduced sensitivity when observing very extended source (which scales as $\theta^{-1}$ with $\theta$ being the extension), observations of Geminga with IACTs are crucial for verifying the MILAGRO claim and helping us to understand the $\gamma$-ray emission from Geminga.

\subsection{PSR B1706$-$44}

Gamma-ray pulsations from PSR~B1706$-$44 were discovered by EGRET; the
observations revealed a triple-peaked pulse profile but no evidence
for unpulsed emission~\citep{egret_psr1706}. More recently, the pulsar
was also detected by \emph{Fermi}/LAT as the bright source
0FGL~J1709.7$-$4428.  The phase-averaged spectra measured by EGRET and
LAT are both well-described by a broken PL (up to 30~GeV, in
the case of the LAT spectrum).  The break energy measured by LAT is
3~GeV, while in deriving the EGRET spectrum, it is fixed at
1~GeV~\citep{egret_psr1706,Bertsch00}. The LAT PL index steepens
from a value of $\sim$1.9 (below 3~GeV) to $\sim$3.3 (above 3~GeV), as shown in
Fig.~\ref{EGR_psr}. This spectrum and the PL spectrum derived
using the method described in Sec.~\ref{Sect:SED_construction} are
both consistent with the photon flux in the 1--100~GeV band reported
in \citet{bsl_lat}.

The discovery of an extended source of VHE emission in the vicinity of
PSR~B1706$-$44 was recently reported by H.E.S.S.\ \citep{hess_1706}.
The TeV source is quite hard ($\Gamma \sim 2.0$), more so than the
high-energy part of the pulsar spectrum.  The VHE $\gamma$-ray
emission is suggested to be related either to a relic PWN of
PSR~B1706$-$44 and/or the SNR~G343.1$-$2.3 \citep{hess_1706}.

\subsection{LS~I $+$61 303}
\label{Sect:LSI}
LS~I $+$61 303 is the first X-ray binary where periodic $\gamma$-ray emission has been detected at both GeV~\citep{lat_LSI61} and TeV energies~\citep{veritas_LSI61,magic_LSI61}. Its SED is shown in Fig.~\ref{XRB}.
The ``bowtie'' looks nicely connected to the measured VHE spectrum, but a cut-off energy at $\sim$6~GeV is reported~\citep{lat_LSI61}. Furthermore, the timing measurements in both the GeV and TeV bands show that the maximum emission occurs at different orbital phases, namely, close to periastron for $<$100~GeV emission and close to apastron for VHE emission. This suggests different emission mechanisms in the two bands, as noted in \citep{lat_LSI61}.

\subsection{Galactic Center Region}
\label{Sect:GC}
The Galactic Center is among the richest and most complex regions in the Galaxy, due to the large number of possible sources and the difficulty of correctly modeling the diffuse emission due to cosmic-ray interaction with the local molecular clouds. This problem is extremely relevant at GeV energies, as demonstrated by EGRET measurements. The discovery of new VHE $\gamma$-ray sources close to the Galactic Center is important for studying the role of diffuse Galactic emission versus the emission from resolved sources in this region~\citep{HowCanFermiHelp_SciNeGHe08}.

One GeV source, 0FGL~J1746.0$-$2900, is detected with a significance of 36$\sigma$ in the neighborhood of the Galactic Center. The best-fit position for 0FGL~J1746.0$-$2900 is R.A.$=17^\mathrm{h} 46^\mathrm{m} 1\farcs4$, Decl.$=-29 \degr 0\arcmin 18\arcsec$ (J2000) with a 95\% C.L. error radius of 4\arcmin~\citep{bsl_lat}. The H.E.S.S. Collaboration also reported a detection of a source towards the Galactic Center, localized at R.A.$=17^\mathrm{h} 45^\mathrm{m} 39\farcs6 \pm 0\farcs4\,\mathrm{(stat)} \pm 0\farcs4\,\mathrm{(sys)}$, Decl.$=-29 \degr 0\arcmin 22\arcsec \pm 6\arcsec\,\mathrm{(stat)} \pm 6\arcsec\,\mathrm{(sys)}$~\citep[J2000;][]{hess_GC_pos}. Based on the procedure described in Sect.~\ref{position_procedure}, the 0FGL~J1746.0$-$2900/HESS~J1745$-$290 pair falls into the category of possibly coincident cases. With better photon statistics, one of the fundamental questions that the LAT can hopefully address is whether or not the GeV and TeV sources are indeed spatially conincident.

The spectra of 0FGL~J1746.0$-$2900 and HESS~J1745$-$290 are shown in Fig.~\ref{GC}. The spectra in the two bands do not appear to be well described by a single PL, and there seems to be an order-of-magnitude drop-off in the energy range $\sim$10--100GeV. Although detailed analysis of the LAT data is out of the scope of this paper, this simple inspection does not indicate that they are from the same emission component (although large uncertainties due to systematics in this region do not permit stronger conclusions at this time). For reference, the broken PL fit of 3EG~J1746$-$2851~\citep{Bertsch00} is also shown in Fig.~\ref{GC}.

\section{Discussion}

In this work, the first comparison of the GeV and VHE $\gamma$-ray sources after the launch of LAT is presented, which takes the advantage of the significantly improved LAT angular resolution and sensitivity compared to EGRET.

Below are a list of preliminary results drawn from this work:

\begin{enumerate}
  \item With the better localization and morphological information of VHE $\gamma$-ray sources compared with 0FGL sources, the nature of the LAT sources may be better revealed through their VHE counterparts. Table~\ref{tab:pos_coincidence1} lists the potential counterparts of some VHE $\gamma$-ray sources which are coincident with 0FGL~sources. For example, HESS~J1804$-$216 may be related to W~30, which may help in understanding the nature of the unidentified source 0FGL~J1805.3$-$2138.
  \item Results of several LAT-detected pulsars show cut-offs at energies $\sim$1--10~GeV, similar to the assessment of \citet{funk08} for EGRET-detected pulsar systems. Therefore, a VHE counterpart ($\sim$0.1--10~TeV) of a LAT pulsar most likely represents the associated PWN, with a shell-type supernova being a viable alternative. This is particularly important for those new pulsars discovered by LAT. The VHE counterparts coincident with the six LAT pulsars may be the associated PWN, although other explanations (e.g. shell-type SNR) are also possible. The question of whether or not typical $\gamma$-ray pulsars are accompanied by VHE-emitting nebulae can be tested by observing them in the VHE domain.
  \item Through broad-band $\gamma$-ray spectra of SNRs, one may in principle distinguish between hadronic and leptonic scenarios. A study of RX~J1713.7$-$3946 using five years of simulated LAT observations  \citep{lat_technical} shows that the energy flux level for the hadronic scenario differs by around a factor of two from that for the leptonic scenario and that a spectral break may be more prominent for the latter. The SNR sample shown in Figs.~\ref{SNR1} and \ref{SNR2} do not seem to support either scenario, although it is too early to draw any conclusion based on the three-month 0FGL dataset. If a hadronic scenario is found to be more viable, this would support the current hypothesis that shell-type SNRs are cosmic-ray sources.
  \item Previous studies did not reveal a strong correlation between the GeV/TeV populations. \citet{reimar_icrc2007} list 16 H.E.S.S. sources without counterparts from the 3EG catalog. Among them, new associations are found in the present study and are presented in Table~\ref{new_associations_0FGL}, thanks mostly to the discovery of new GeV sources with LAT. \citet{reimar_icrc2007} also present 11 sources in the 3EG catalog without H.E.S.S. counterpart. Among them, 0FGL~J1709.7$-$4428 \citep[the 0FGL counterpart of 3EG~J1710$-$4439;][]{bsl_lat} is now found to be associated with HESS~J1708$-$443, a source discovery reported in~\citet{hess_1706}.
  \item All spatially coincident GeV and TeV pairs during the EGRET era are essentially consistent with one single spectral component \citep[See Figs. 4--6 in][]{funk08}. With the significantly enhanced sensitivity of LAT, new relations between the GeV and TeV spectra are apparent in the SEDs. The SNR candidate 0FGL~J1834.4$-$0841 and the unidentified 0FGL~J1805.3$-$2138 (and their likely VHE counterparts) represent the first examples for which the GeV/TeV spectrum cannot be treated as a single emission component. A similar conclusion is reached by~\citet{lat_LSI61} for a HMXB (LS~I~$+$61~303), based on the light curves and spectral incompatibility of this source in the two bands.
  \item \citet{milagro_bsl} consider a probability that many unidentified LAT sources are extragalactic, so as to explain the low rate of finding coincident MILAGRO emission among the unidentified LAT sources. This idea might also explain the non-detection of VHE counterparts of a majority of the unidentified LAT sources. On the other hand, the extended nature of all the five spatially coincident cases (HESS~J1023$-$575, HESS~J1804$-$216, HESS~J1841$-$055, HESS~J1843$-$033, HESS~J1848$-$018; if proved to be real association) would exclude an extragalactic origin of the corresponding LAT sources.
  \item Although VHE observations only cover a small part of the whole sky, they do cover the majority of the inner Galaxy, e.g., the H.E.S.S. telescopes have surveyed the region of $l=-$85$\degr$ to 60$\degr$, $b=-$3$\degr$ to 3$\degr$~\citep{hess_survey09}.
In this region, there are 41 \emph{Fermi} bright sources. Among them, 16 are found coincident with a VHE counterpart. This fraction ($\sim$2/5) is higher than that for EGRET where about 1/4 of the EGRET sources (in a smaller region) are found to have a coincident VHE counterpart~\citep{funk08}. Moreover, the number rises to 21 (out of 41) if possibly coincident cases are included and the fraction becomes 50\%. The LAT radii of the 95\% confidence region are in general much smaller than the EGRET error boxes, which further strengthens the case of a higher fraction for LAT. Even though the VHE extension is taken into account in this study \citep[but not in][]{funk08}, a typical extension is of the same order as a LAT positional uncertainty. A breakdown of the number of coincidence cases for each source population in the above-defined region of the H.E.S.S. Galactic Plane Survey is shown in Table~\ref{breakdown}.
\end{enumerate}

\section{Conclusion}

In this work, we search for VHE counterparts of each Galactic GeV source in the 0FGL catalog~\citet{bsl_lat}, based on spatial coincidence. This study benefits significantly from the increased LAT angular resolution and its better sensitivity over previous instruments.

Compared to the EGRET era, not only are there more coincident sources (improvement in quantity), but improvements in quality start to emerge. With the much better sensitivity of LAT, weaker sources are detected which were unknown in the EGRET era. New relations between the GeV and TeV spectra are revealed. A single spectral component is unable to describe some sources detected at both GeV and TeV energies. Two spectral components may be needed in these cases to accommodate the SEDs, where the VHE flux is higher than a PL extrapolation from GeV energies.

A high fraction of \emph{Fermi} bright sources are found to be spatially coincident with a VHE $\gamma$-ray source. This shows that there exists a common GeV/TeV source population, a conclusion that is in stark disagreement with that of~\citet{funk08}.

\begin{acknowledgements}
We thank Emma O\~na-Wilhelmi and Werner Hofmann for useful discussion.

\end{acknowledgements}

   \begin{figure*}
   \begin{minipage}[t]{160mm}
     \centering
    \includegraphics[width=\columnwidth]{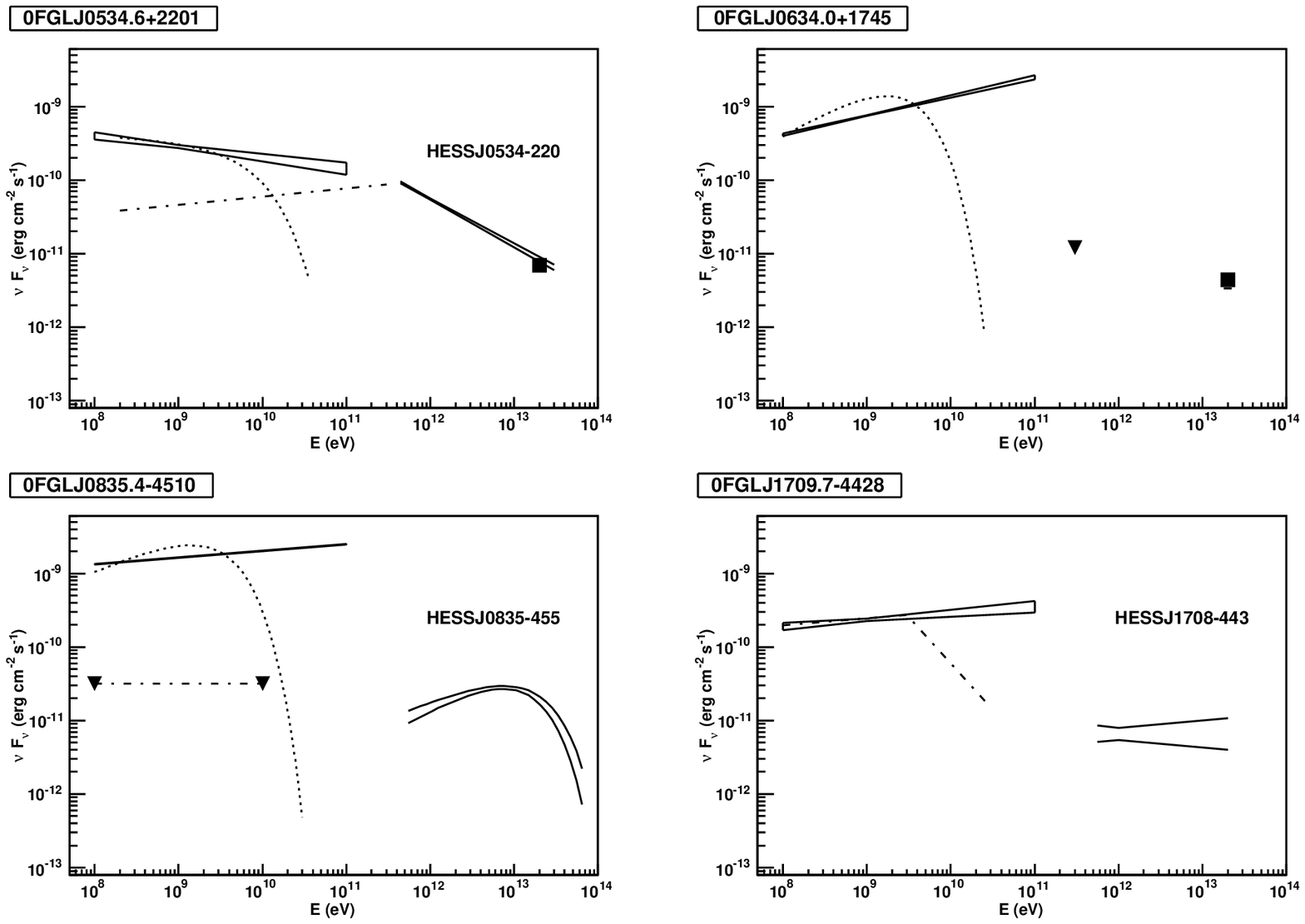}
      \caption{MeV to TeV spectra of four EGRET pulsars and their proposed nebulae. \emph{Upper left}: Crab (0FGL~J0534.6$+$2201). The pulsar (dotted line) and nebula (dashed-dotted line) spectral components are those reported in~\citet{lat_crab}. The VHE spectra are taken from~\citet{hess_crab}, and the MILAGRO measurement at 20~TeV is shown~\citep{milagro_GPS}. \emph{Upper right}: Geminga (0FGL~J0634.0$+$1745). The pulsar spectrum (dotted line) is that reported in~\citet{lat_geminga}. The triangle denotes the upper limit reported in~\citet{veritas_geminga}, and the MILAGRO measurement at 20~TeV is also indicated~\citep{milagro_GPS}. \emph{Lower left}: Vela (0FGL~J0835.4$-$4510). The dotted line represents the Vela spectrum as shown in~\citet{lat_velapsr}, while the nebula component is constrained by the two triangles joined by the dashed-dotted line. The curved VHE spectrum is taken from~\citet{hess_velaX}. \emph{Lower right}: PSR~B1706$-$44 (0FGL~J1709.7$-$4428). The dashed-dotted line denotes the two PL model spectrum derived in \citet{lat_egr_psr}. Both LAT energy spectra (though different above 3~GeV) are consistent with the photon flux in the 1--100~GeV band of this source~\citep{bsl_lat}. The VHE spectrum is taken from~\citet{hess_1706}.}
      \label{EGR_psr}
      \end{minipage}
   \end{figure*}

   \begin{figure*}
   \begin{minipage}[t]{160mm}
     \centering
    \includegraphics[width=\columnwidth]{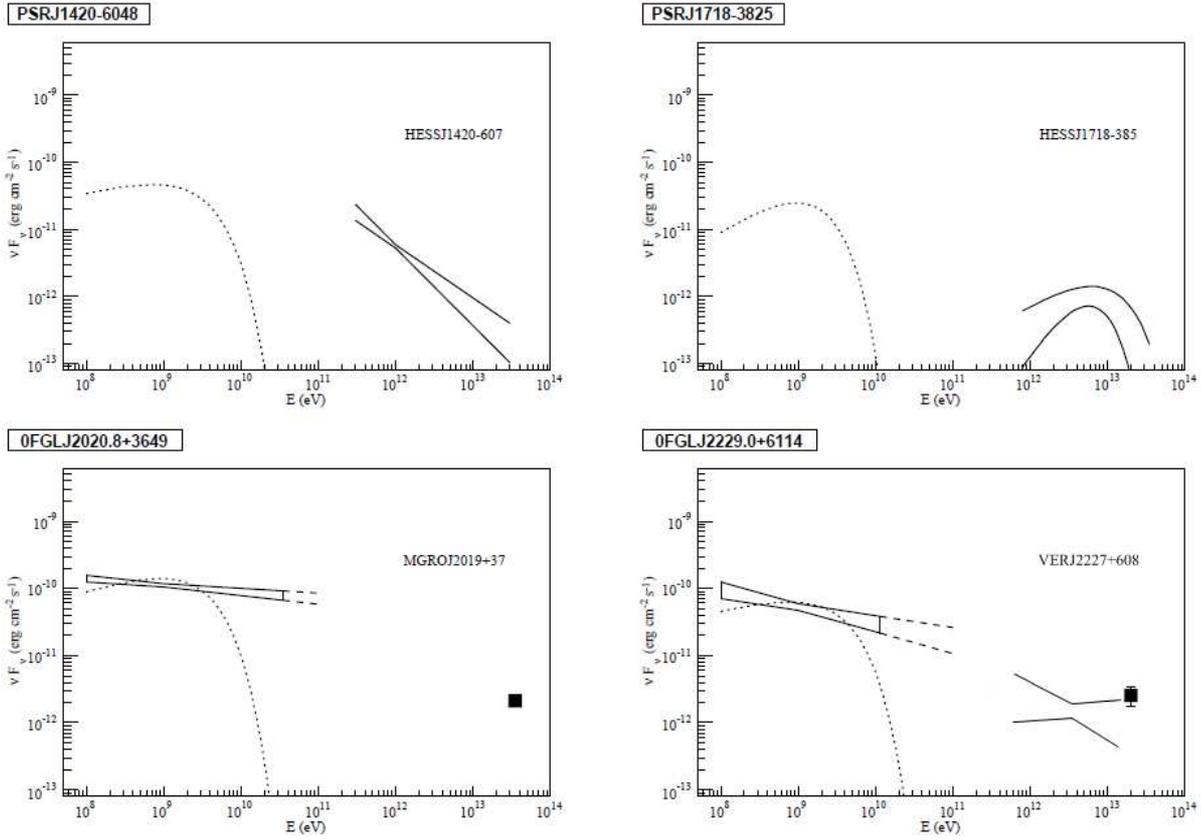}
      \caption{MeV to TeV spectra of the four radio pulsars first detected in $\gamma$-rays by LAT. The spectra below 300~GeV are taken from~\citet{lat_psr_cat}. \emph{Upper left}: The VHE spectrum is taken from~\citet{hess_kookaburra}. \emph{Upper right}: The VHE spectrum presented in~\citet{hess_psr_1718_1809} is shown. The two curves represent the upper and lower limits of the spectrum, taking measurement errors into account. \emph{Lower left}: The flux at 35~TeV is taken from \citet{milagro_bsl}. \emph{Lower right}: The flux at 20~TeV is taken from \citet{milagro_GPS} and the VHE spectrum is taken from \citet{veritas_2227}.}
      \label{radio_psr}
      \end{minipage}
   \end{figure*}

   \begin{figure*}
      \begin{minipage}[t]{160mm}
     \centering
    \includegraphics[width=\columnwidth]{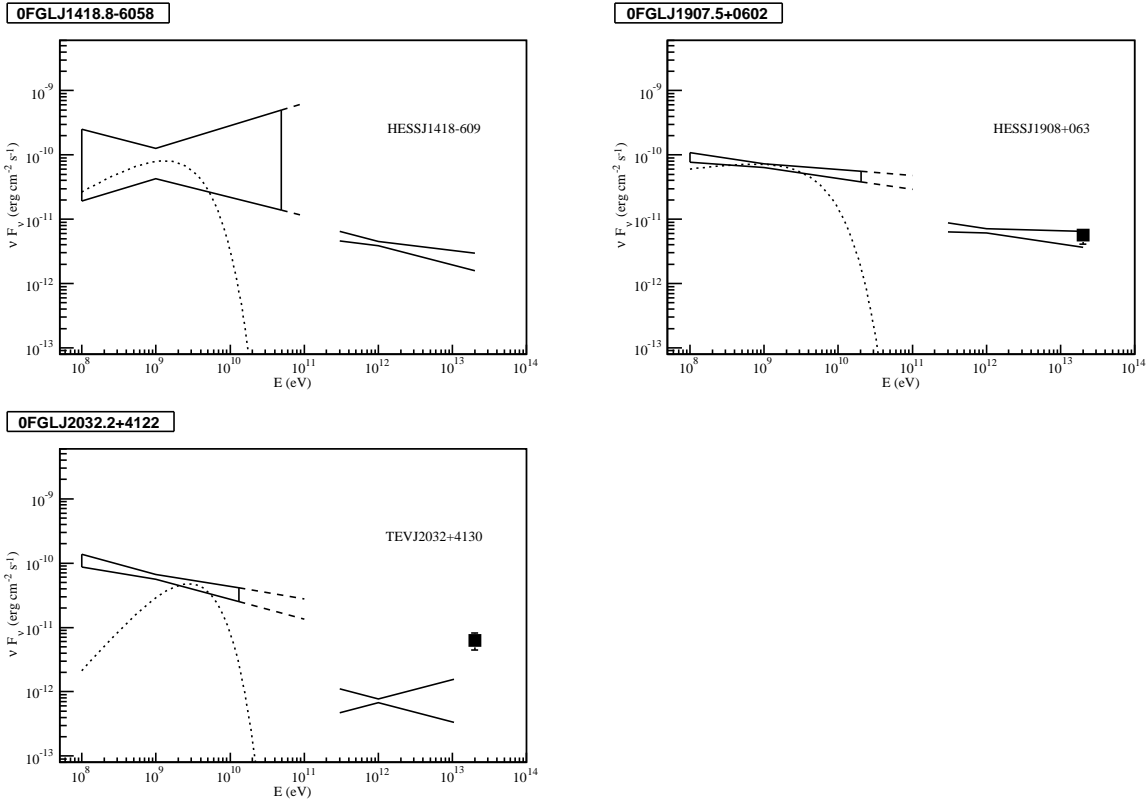}
      \caption{MeV to TeV spectra of the three new pulsars discovered in a blind search for pulsation in the LAT data. The spectra below 300~GeV are taken from~\citet{lat_psr_cat}. \emph{Upper left}: The VHE spectrum is taken from~\citet{hess_kookaburra}. \emph{Upper right}: The VHE spectrum presented in~\citet{icrc07_1908+063} is shown, together with the coincident MILAGRO source flux at 20~TeV~\citep{milagro_GPS}. \emph{Lower left}: The VHE spectrum is that presented in~\citet{hegra_TeV2032}, while the MILAGRO flux at 20~TeV is also shown~\citep{milagro_GPS}.}
         \label{LAT_psr}
   \end{minipage}
   \end{figure*}

   \begin{figure*}
   \begin{minipage}[t]{160mm}
     \centering
    \includegraphics[width=\columnwidth]{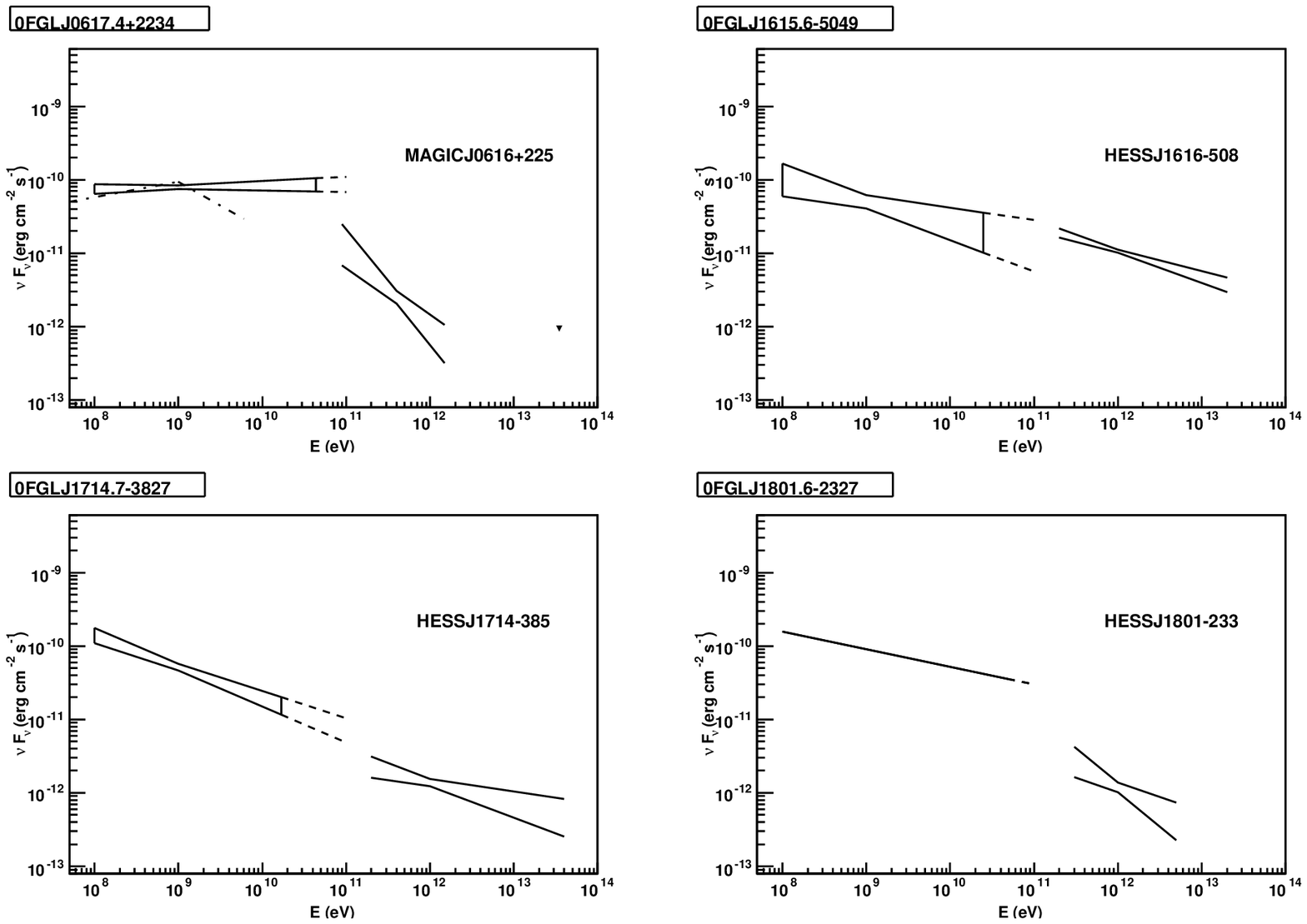}
      \caption{MeV to TeV spectra of four SNR/PWN candidate 0FGL sources. Spectra at $>$100~GeV energies are taken from \citet{magic_ic443} (MAGIC~J0616$+$225), \citet{hess_survey} (HESS~J1616$-$508), \citet{CTB_37A} (HESS~J1714$-$385), and \citet{hess_1801-233} (HESS~J1801$-$233). The broken PL spectrum (dashed-dotted line) derived for 3EG~J0617$+$2238 is taken from \citet{Bertsch00}. The flux at 35~TeV at the position of 0FGL~J0617.4$+$2234 is that given in~\citet{milagro_bsl}.}
      \label{SNR1}
      \end{minipage}
   \end{figure*}

   \begin{figure*}
      \begin{minipage}[t]{160mm}
     \centering
    \includegraphics[width=\columnwidth]{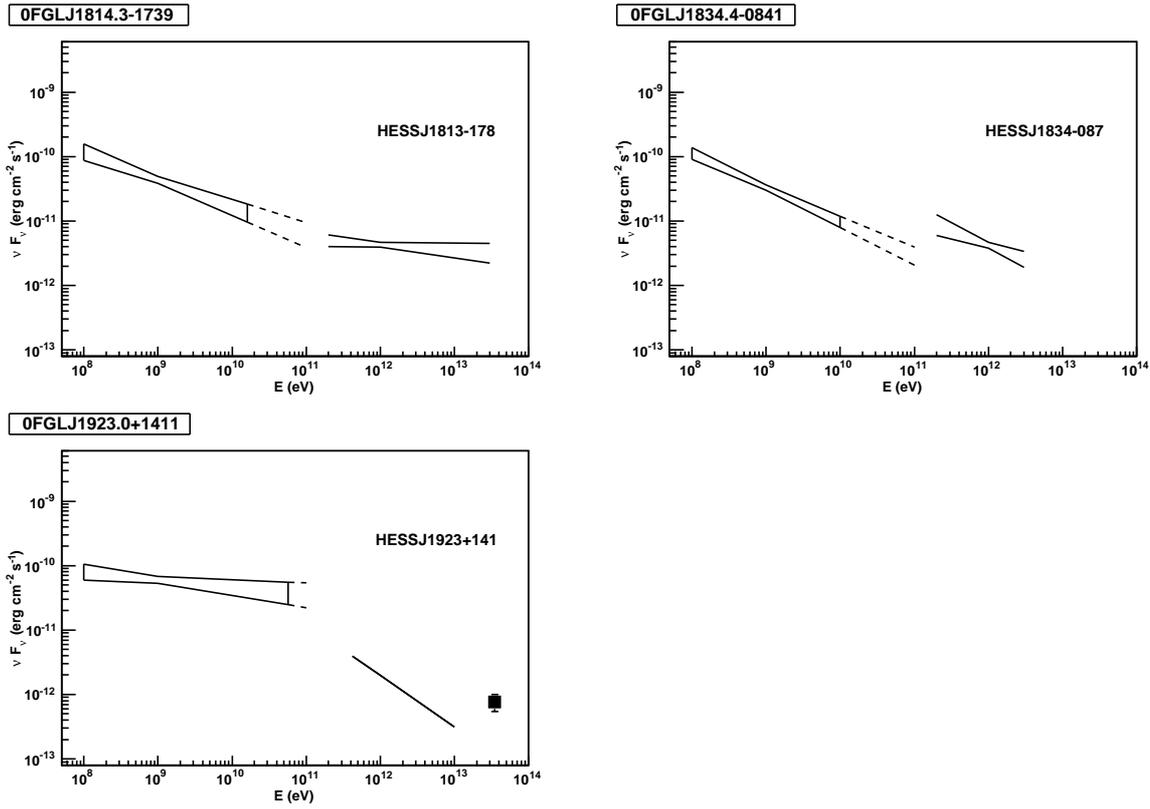}
      \caption{MeV to TeV spectra of three SNR/PWN candidate 0FGL sources. Spectra at $>$100~GeV energies are taken from \citet{hess_survey} (HESS~J1813$-$178 and HESS~J1834$-$087). For HESS~J1923$+$141, an assumed photon index of 2.8 is used in deriving the spectrum using the flux given in~\citet{hess_w51}, and the flux at 35~TeV is that given in~\citet{milagro_bsl}. There is evidence of a steepening above several GeV~\citep{lat_W51C}.}
      \label{SNR2}
   \end{minipage}
   \end{figure*}

   \begin{figure*}
   \begin{minipage}[t]{160mm}
     \centering
    \includegraphics[width=\columnwidth]{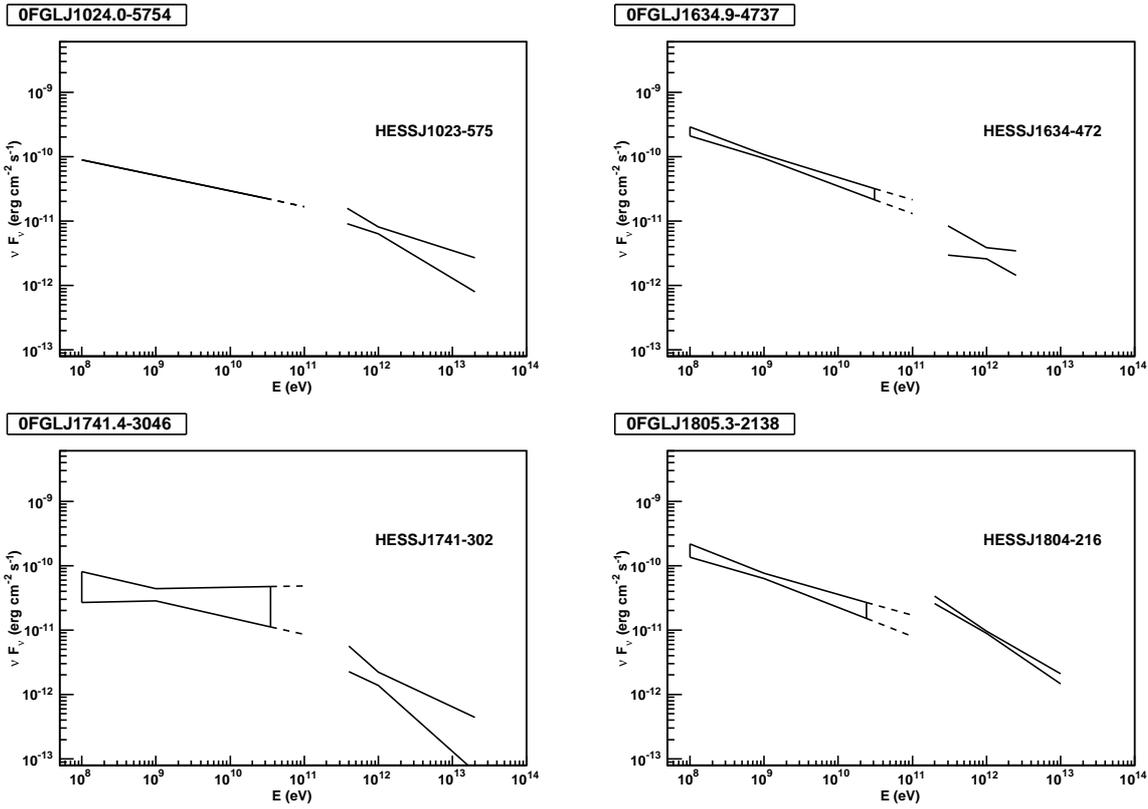}
      \caption{MeV to TeV spectra of four unidentified 0FGL sources. Spectra at $>$100~GeV energies are taken from \citet{hess_westerlund2_1023} (HESS~J1023$-$575), \citet{hess_survey} (HESS~J1634$-$472 and HESS~J1804$-$216), and \citet{hess_1741_HDGS} (HESS~J1741$-$302).}
      \label{Unid1}
      \end{minipage}
   \end{figure*}

   \begin{figure*}
      \begin{minipage}[t]{160mm}
     \centering
    \includegraphics[width=\columnwidth]{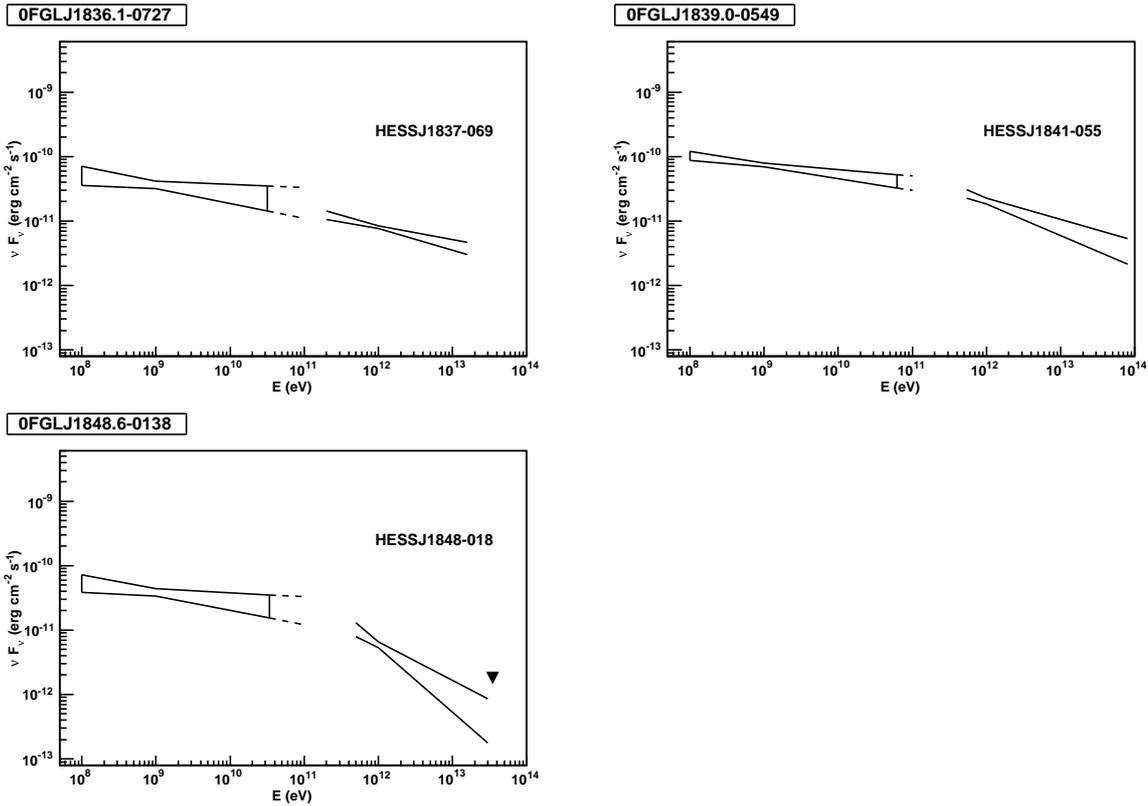}
      \caption{MeV to TeV spectra of three unidentified 0FGL sources. Spectra at $>$100~GeV energies are taken from \citet{hess_survey} (HESS~J1837$-$069), \citet{hess_dark} (HESS~J1841$-$055), and \citet{hess_1848_HDGS} (HESS~J1848$-$018). The flux at 35~TeV at the position of 0FGL~J1848.6$-$0138 is that given in~\citet{milagro_bsl}.}
      \label{Unid2}
   \end{minipage}
   \end{figure*}

   \begin{figure*}
      \begin{minipage}[t]{160mm}
     \centering
    \includegraphics[width=\columnwidth,height=55mm]{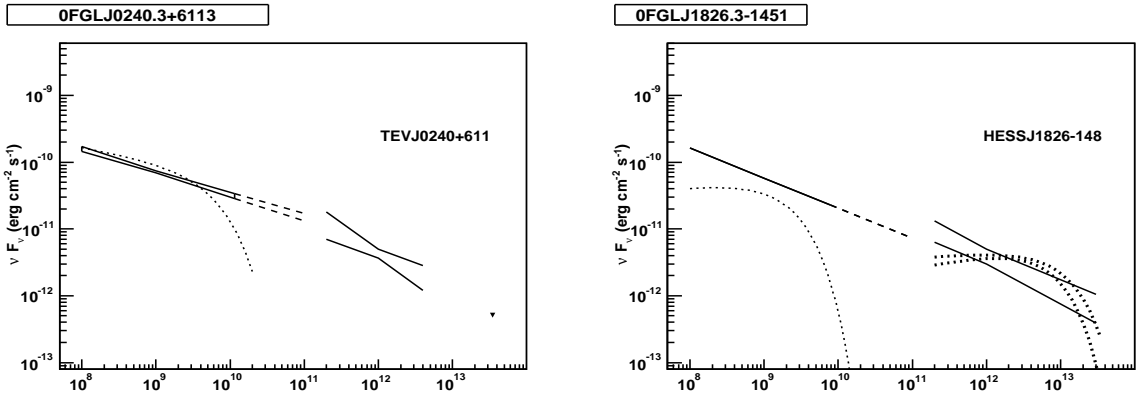}
      \caption{MeV to TeV spectra of the two X-ray binaries in the 0FGL catalog. The phase-averaged exponential cut-off spectrum in the GeV range of LS~I~$+$61~303 (\emph{left}) is taken from \citet{lat_LSI61}. That of LS~5039 is taken from~\citet{lat_ls5039}. Spectra at $>$100~GeV energies are taken from \citet{magic_LSI61} (for a partial phase of LS~I~$+$61~303 during which VHE emission is detected) and \citet{hess_LS5039} (for two phases of LS~5039). The flux at 35~TeV for LS~I~$+$61~303 is that given in~\citet{milagro_bsl}.}
         \label{XRB}
   \end{minipage}
   \end{figure*}

   \begin{figure*}
      \begin{minipage}[t]{160mm}
     \centering
    \includegraphics[width=.499\columnwidth]{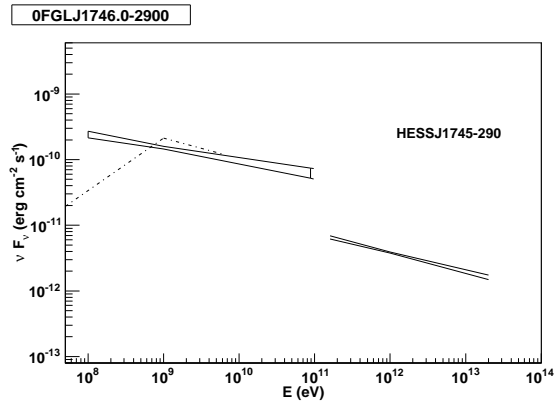}
      \caption{MeV to TeV spectrum of the Galactic Center. The VHE spectrum is taken from \citet{hess_GC_DarkMatter} while the broken PL spectrum (dashed-dotted line) derived for 3EG~J1746$-$2851 is taken from \citet{Bertsch00}.}
         \label{GC}
   \end{minipage}
   \end{figure*}

   \begin{figure*}
      \begin{minipage}[t]{130mm}
     \centering
    \includegraphics[width=\columnwidth]{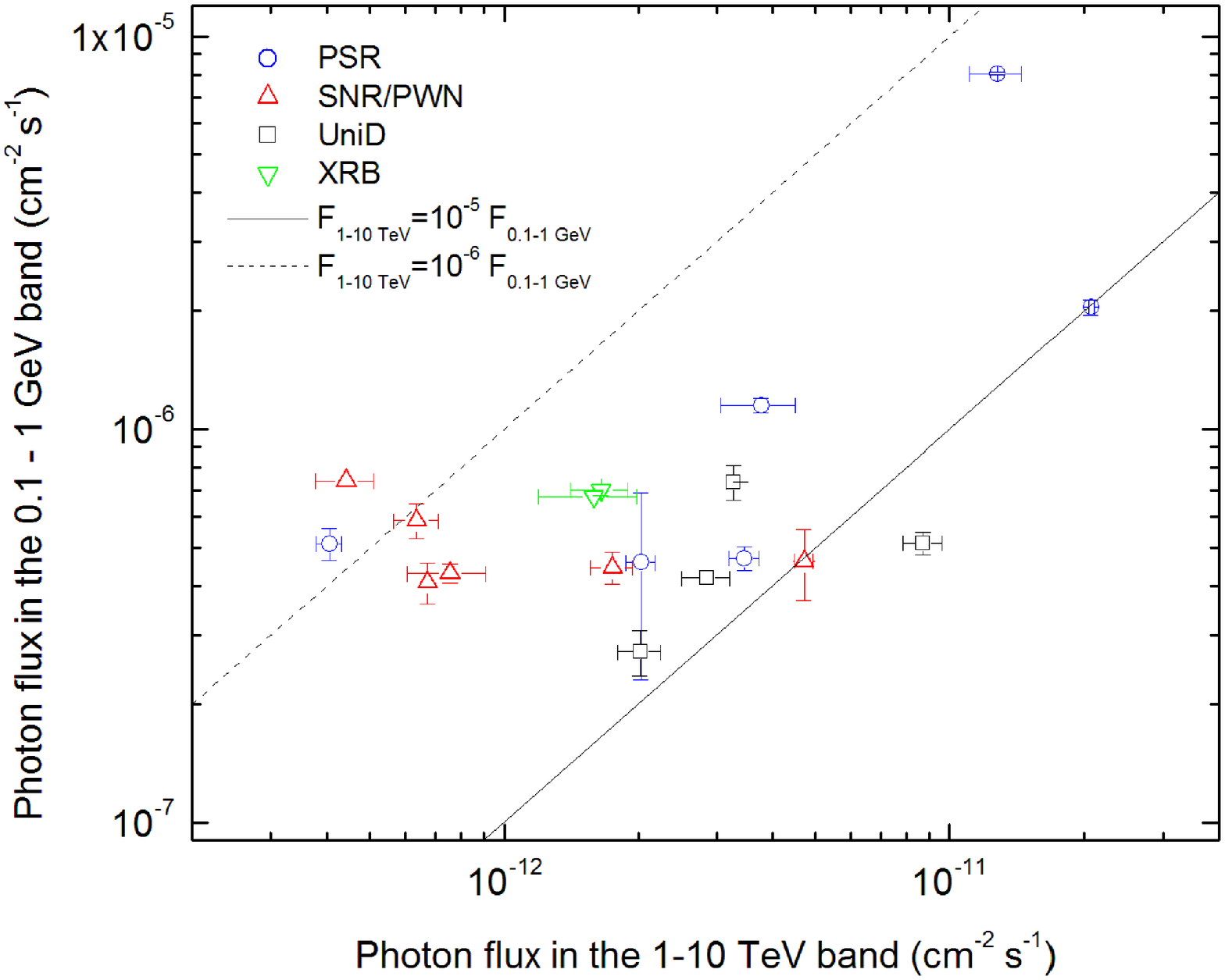}
      \caption{Photon flux in the 100~MeV -- 1~GeV band versus that in the 1--10~TeV band for coincident GeV/TeV sources.}
      \label{F23}
   \end{minipage}
   \end{figure*}

   \begin{figure*}
      \begin{minipage}[t]{130mm}
     \centering
       \includegraphics[width=\columnwidth]{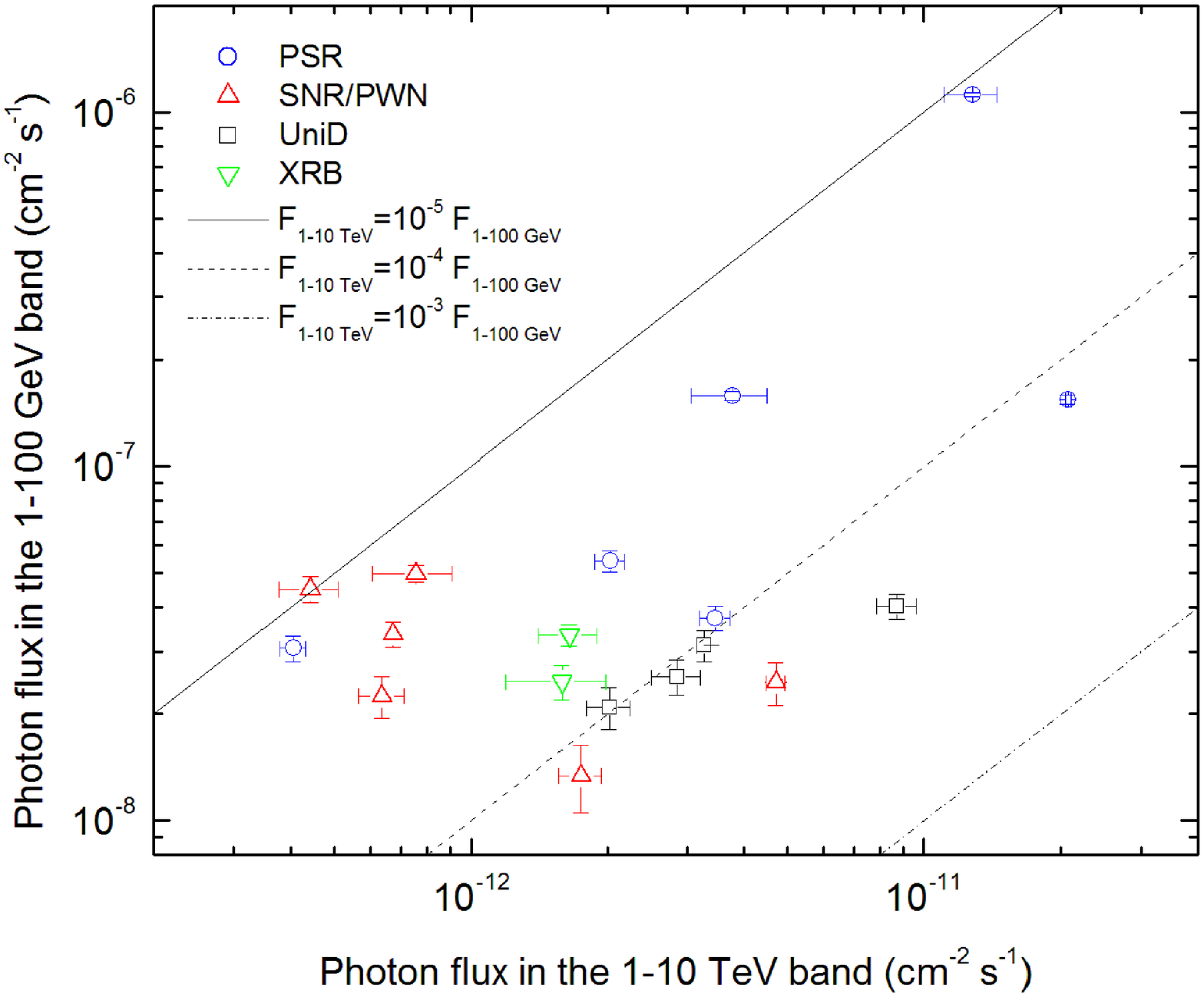}
      \caption{Photon flux in the 1--100~GeV band versus that in the 1--10~TeV band for coincident GeV/TeV sources.}
      \label{F35}
   \end{minipage}
   \end{figure*}

   \begin{figure*}
      \begin{minipage}[t]{130mm}
     \centering
       \includegraphics[width=\columnwidth]{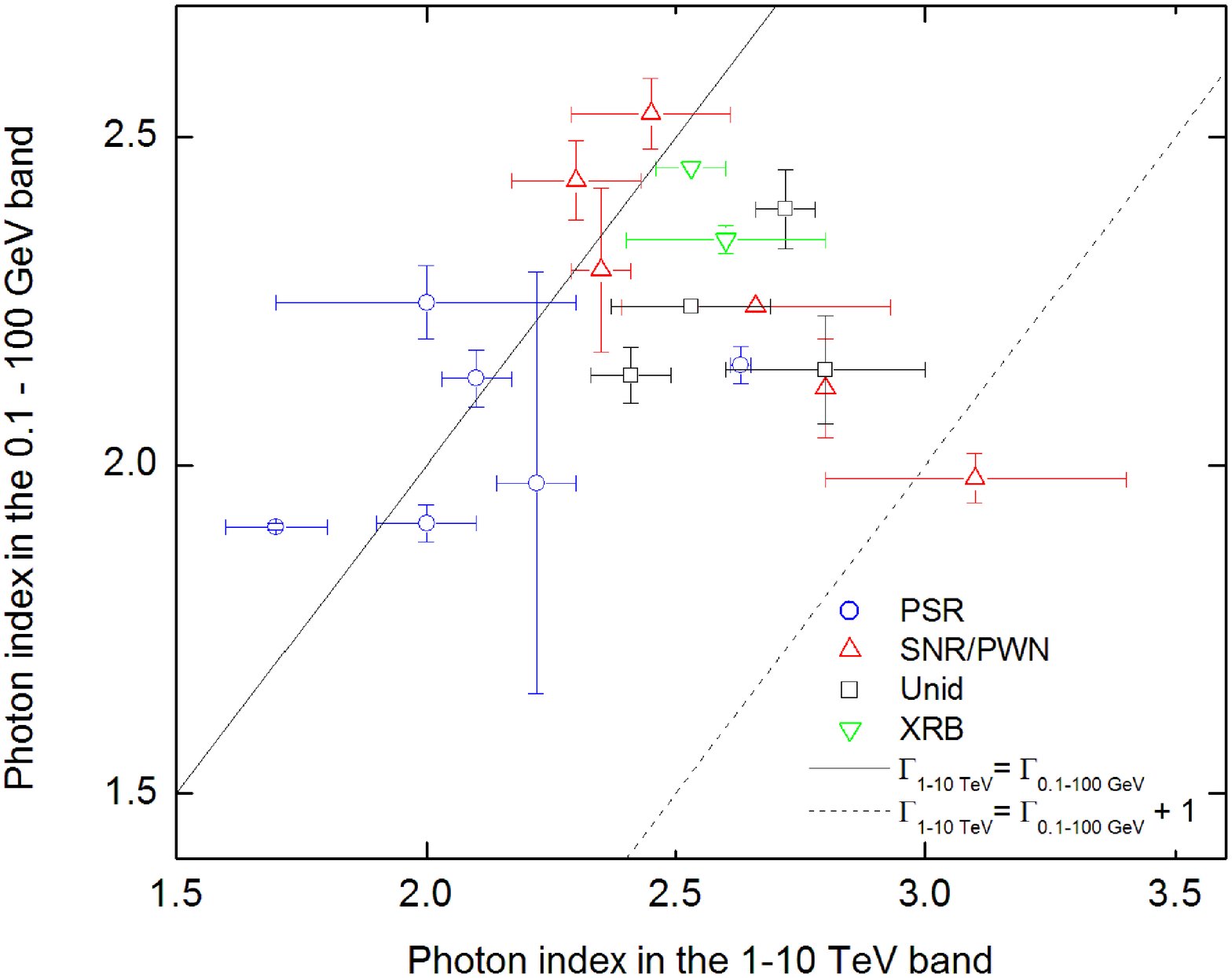}
      \caption{Photon index in the 0.1--100~GeV band (derived according to Sect.~\ref{Sect:SED_construction}) versus photon index in the 1--10~TeV band, for coincident GeV/TeV sources. 0FGL~J1923.0$+$1411/HESS~J1923$+$141 is not included since its VHE photon index is not known.}
      \label{Gamma}
   \end{minipage}
   \end{figure*}

\begin{landscape}
\begin{table}
\begin{minipage}[t]{180mm}
\caption{0FGL sources and LAT pulsars with spatially coincident VHE counterpart}
\label{tab:pos_coincidence1}
\centering
\renewcommand{\footnoterule}{}  
\begin{tabular}{cccccc|cccccc}
\hline\hline
     LAT source        & association\tablefootmark{a} & class\tablefootmark{b} & $l$ & $b$ & error\tablefootmark{c} & VHE $\gamma$-ray source & association\tablefootmark{d} & $l$ & $b$ & extension\tablefootmark{e} & references \\
               &  & & ($\degr$) & ($\degr$) & ($\degr$) & & & ($\degr$) & ($\degr$) & ($\degr$) & \\
    \hline
    0FGL~J0534.6$+$2201 & Crab    & PSR & 184.56 & $-$5.76 & 0.05 & HESS~J0534$+$220 & Crab nebula & 184.56 & $-$5.78 & PS & 1 \\
    0FGL~J0835.4$-$4510 & Vela    & PSR & 263.56 & $-$2.77 & 0.04 & HESS~J0835$-$455 & Vela~X & 263.86 & $-$3.09 & 0.43 & 2 \\
    0FGL~J1418.8$-$6058 &         & PSR & 313.34 & 0.11 & 0.07 & HESS~J1418$-$609 & G313.3$+$0.1 (Rabbit) & 313.25 & 0.15 & 0.06 & 3 \\
    PSR~J1420$-$6048    &         & PSR & 313.5 & 0.2 & PS & HESS~J1420$-$607 & PSR~J1420$-$6048 & 313.56 & 0.27 & 0.07 & 3 \\
    0FGL~J1709.7$-$4428 & PSR~B1706$-$44 & PSR & 343.11 & $-$2.68 & 0.05 & HESS~J1708$-$443\tablefootmark{f} & PSR~B1706$-$44? & 343.04 & $-$2.38 & 0.29 & 4 \\
    PSR~J1718$-$3825    &         & PSR & 349.0 & $-$0.4 & PS & HESS~J1718-385 & & 348.83 & $-$0.49 & 0.015 & 5 \\
    0FGL~J1907.5$+$0602 &         & PSR & 40.14 & $-$0.82 & 0.08 & HESS~J1908$+$063 & & 40.39 & $-$0.79 & 0.34 & 6 \\
    0FGL~J2032.2$+$4122 &         & PSR & 80.16 & 0.98 & 0.09 & TeV~J2032$+$4130 & & 80.23 & 1.10 & 0.10 & 7 \\
    0FGL~J0617.4$+$2234 &         & SNR/PWN & 189.08 & 3.07 & 0.06 & VER~J0616.9$+$2230 & IC~443 & 189.08 & 2.92 & 0.16 & 8 \\
    0FGL~J1615.6$-$5049 &         & SNR/PWN & 332.35 & $-$0.01 & 0.23 & HESS~J1616$-$508 & PSR~J1617$-$5055? & 332.39 & -0.14 & 0.14 & 9 \\
    0FGL~J1648.1$-$4606 &         & SNR/PWN & 339.47 & $-$0.71 & 0.18 & Westerlund 1 region\tablefootmark{f,g} & & 339.55 & $-$0.40 & $\sim$0.9\tablefootmark{h} & 10 \\
    0FGL~J1714.7$-$3827 &         & SNR/PWN & 348.53 & 0.1 & 0.13 & HESS~J1714$-$385 & CTB~37A & 348.39 & 0.11 & 0.07 & 11 \\
    0FGL~J1801.6$-$2327 &         & SNR/PWN & 6.54 & $-$0.31 & 0.11 & HESS~J1801$-$233 & W~28 & 6.66 & $-$0.27 & 0.17 & 12 \\
    0FGL~J1834.4$-$0841 &         & SNR/PWN & 23.27 & $-$0.22 & 0.1 & HESS~J1834$-$087 & W~41 & 23.24 & $-$0.32 & 0.09 & 9 \\
    0FGL~J1923.0$+$1411 & W~51C\tablefootmark{i} & SNR  & 49.13 & $-$0.4 & 0.08 & HESS~J1923$+$141\tablefootmark{f} & W~51 & 49.14 & $-$0.6 & $\sim$0.15\tablefootmark{j} & 13 \\
    0FGL~J1024.0$-$5754 &         & Unid & 284.35 & $-$0.45 & 0.11 & HESS~J1023$-$575 & Westerlund 2? & 284.19 & $-$0.39 & 0.18 & 14 \\
    0FGL~J1805.3$-$2138 &         & Unid & 8.54 & $-$0.17 & 0.19 & HESS~J1804$-$216 & W~30/PSR~J1803$-$2137? & 8.40 & -0.03 & 0.20 & 9 \\
    0FGL~J1839.0$-$0549 &         & Unid & 26.34 & 0.08 & 0.12 & HESS~J1841$-$055 & & 26.8 & $-$0.2 & 0.42 & 15 \\
    0FGL~J1844.1$-$0335 &         & Unid & 28.91 & $-$0.02 & 0.15 & HESS~J1843$-$033\tablefootmark{f} & & $\sim$29.08 & $\sim$0.16 & $\sim$0.2\tablefootmark{j} & 16 \\
    0FGL~J1848.6$-$0138 &         & Unid & 31.15 & $-$0.12 & 0.16 & HESS~J1848$-$018\tablefootmark{f} & & 30.98 & $-$0.16 & 0.32 & 17 \\
    0FGL~J0240.3$+$6113 & LS~I~$+$61~303 & HMXB & 135.66 & 1.08 & 0.07 & VER~J0240$+$612 & LS~I~$+$61~303 & 135.70 & 1.08 & PS & 18 \\
    0FGL~J1826.3$-$1451 & LS 5039      & HMXB & 16.89 & $-$1.32 & 0.11 & HESS~J1826$-$148 & LS~5039 & 16.90 & $-$1.28 & PS & 19 \\
    \hline
\end{tabular}
\tablefoot{
The top panel shows likely members of Pismis~11. The second panel contains likely
members of Alicante~5. The bottom panel displays stars outside the clusters.\\
\tablefoottext{a}{based on timing information}
\tablefoottext{b}{Source class according to~\citet{bsl_lat}. PSR: pulsar; SNR/PWN: supernova remnant/PWN; HMXB: high-mass X-ray binary; Unid: unidentified sources. The classification of those sources as SNR/PWN is based on spatial coincidence only.}
\tablefoottext{c}{95\% positional error}
\tablefoottext{d}{based on spatial coincidence only}
\tablefoottext{e}{An entry of ``PS'' indicates that the source is point-like with respect to the point spread function of the respective instrument.}
\tablefoottext{f}{These recent source discoveries are preliminary and they have been published in the referenced conference proceedings only.}
\tablefoottext{g}{The VHE emission has been detected by H.E.S.S. towards the direction of the massive stellar cluster. The coordinates refer to the nominal position of Westerlund~1.}
\tablefoottext{h}{The extent of the source is not clear. The given value is estimated from the radial profile shown in Fig. 4 of the corresponding reference.}
\tablefoottext{i}{The association is based on a morphological study~\citep{lat_W51C}.}
\tablefoottext{j}{The extent of the source is not clear. The given value is estimated from the sky excess map in the corresponding reference.}
}
\tablebib{
(1)~\citet{hess_crab}; (2)~\citet{hess_velaX}; (3)~\citet{hess_kookaburra}; (4)~\citet{hess_1706}; (5)~\citet{hess_psr_1718_1809}; (6)~\citet{hess_1908+063}; (7)~\citet{hegra_TeV2032}; (8)~\citet{veritas_ic443};
(9)~\citet{hess_survey}; (10)~\citet{hess_westerlund1_1648}; (11)~\citet{CTB_37A}; (12)~\citet{hess_1801-233}; (13)~\citet{hess_w51}; (14)~\citet{hess_westerlund2_1023}; (15)~\citet{hess_dark}; (16)~\citet{hess_survey_2007}; (17)~\citet{hess_1848_HDGS}; (18)~\citet{veritas_LSI61}; (19)~\citet{hess_LS5039}.
}
\end{minipage}
\end{table}


\begin{table}
\begin{minipage}[t]{\columnwidth}
\caption{0FGL sources with coincident MILAGRO source, but without plausible coincident reported VHE $\gamma$-ray sources. See Table~\ref{tab:pos_coincidence1} for the nomenclature.}
\label{tab:pos_coincidence2}
\centering
\renewcommand{\footnoterule}{}  
\begin{tabular}{lcccc|ccccc}
\hline\hline
    LAT source        & Class & $l$ & $b$ & error & MILAGRO source & $l$ & $b$ & extension & references \\
               & & ($\degr$) & ($\degr$) & ($\degr$) & ($\degr$) & ($\degr$) & ($\degr$) &  \\
    \hline
    0FGL~J0634.0$+$1745 & PSR & 195.16 & 4.29 & 0.04   & MGRO~C3       & 195.3 & 3.8 & 1.3 & \citet{milagro_bsl} \\
    0FGL~J2020.8$+$3649 & PSR & 75.182 & 0.131 & 0.060 & MGRO~J2019$+$37 & 74.8 & 0.4 & $\sim$0.1 & \citet{milagro_bsl} \\
    \hline
\end{tabular}
\end{minipage}
\end{table}
\end{landscape}

\begin{landscape}
\begin{table}
\begin{minipage}[t]{180mm}
\caption{0FGL sources with a possibly coincident VHE $\gamma$-ray source. See Table~\ref{tab:pos_coincidence1} for the nomenclature.}
\label{tab:pos_coincidence3}
\centering
\renewcommand{\footnoterule}{}  
\begin{tabular}{ccccc|cccccc}
\hline\hline
    LAT source         & class & $l$ & $b$ & error & VHE $\gamma$-ray source & association & $l$ & $b$ & extension & references \\
               &   & ($\degr$) & ($\degr$) & ($\degr$) & & & ($\degr$) & ($\degr$) & ($\degr$) & \\
    \hline
    0FGL~J1814.3$-$1739 & SNR/PWN & 13.05 & $-$0.09 & 0.19 & HESS~J1813$-$178 & G12.8$-$0.2/AX~J1813$-$178 & 12.81 & -0.03 & 0.04 & 1 \\
    0FGL~J1634.9$-$4737 & Unid & 336.84 & $-$0.03 & 0.08 & HESS~J1634$-$472 & & 337.11 & 0.22 & 0.11 & 1 \\
    0FGL~J1741.4$-$3046 & Unid & 357.96 & $-$0.19 & 0.2 & HESS~J1741$-$302\tablefootmark{a} & & 358.4 & 0.01 & ? & 2 \\
    0FGL~J1746.0$-$2900 & Unid & 359.99 & $-$0.11 & 0.07 & HESS~J1745$-$290 & Sgr~A*/G359.95$-$0.04 & 359.94 & $-$0.04 & PS & 3 \\
    0FGL~J1836.1$-$0727 & Unid & 24.56 & $-$0.03 & 0.22 & HESS~J1837$-$069 & & 25.18 & $-$0.12 & 7$\farcm$2$\times$3$\arcmin$ & 1 \\
    0FGL~J2021.5$+$4026 &  PSR & 78.23 & 2.07 & 0.05 & VER~J2019$+$407\tablefootmark{a}  & $\gamma$ Cygni SNR? & 78.33 & 2.54 & 0.16$\times$0.11 & 4 \\
    0FGL~J2229.0$+$6114 &  PSR & 106.64 & 2.96 & 0.08 & VER~J2227$+$608  & & 106.35 & 2.71 & 0.27$\times$0.18 & 5 \\
    \hline
\end{tabular}
\tablefoot{
\tablefoottext{a}{These recent source discoveries are preliminary and they have been published in the referenced conference proceedings only.}
}
\tablebib{
(1)~\citet{hess_survey}; (2)~\citet{hess_1741_HDGS}; (3)~\citet{hess_GC_pos}; (4)~\citet{VERITAS_cygnus_survey09}; (5)~\citet{veritas_2227}.
}
\end{minipage}
\end{table}



\begin{table}
\begin{minipage}[t]{\columnwidth}
\centering
\caption{H.E.S.S. sources which have a coincident 0FGL source but do not have a 3EG counterpart as in \citet{reimar_icrc2007}}
\label{new_associations_0FGL}
\centering
\renewcommand{\footnoterule}{}  
\begin{tabular}{lcc}
\hline\hline
H.E.S.S. sources  & 0FGL sources & coincidence level  \\
\hline
    HESS~J1616$-$508 & 0FGL~J1615.6$-$5049 & {\it Y}  \\
    HESS~J1634$-$472 & 0FGL~J1634.9$-$4737 & {\it P}  \\
    HESS~J1745$-$290 & 0FGL~J1746.0$-$2900 & {\it P}  \\
    HESS~J1804$-$216 & 0FGL~J1805.3$-$2138 & {\it Y}  \\
    HESS~J1834$-$087 & 0FGL~J1834.4$-$0841 & {\it Y}  \\
    HESS~J1837$-$069 & 0FGL~J1836.1$-$0727 & {\it P}  \\
\hline
\end{tabular}
\end{minipage}
\end{table}


\begin{table}
\begin{minipage}[t]{\columnwidth}
\centering
\caption{Number of coincidence cases for each source population (excluding extragalactic sources) in the region $l=-$85$\degr$ to 60$\degr$, $b=-$3$\degr$ to 3$\degr$.}
\label{breakdown}
\centering
\renewcommand{\footnoterule}{}  
\begin{tabular}{lcc}
\hline\hline
                &              & spatially    \\
LAT Source class  & 0FGL sources & coincident cases\tablefootmark{a} \\
\hline
pulsars              & 10      &  4      \\
SNR/PWN candidates   & 11      &  6 (7)   \\
Unidentified sources & 19      &  5 (9)   \\
\hline
Total\tablefootmark{b} & 41      &  16 (21) \\
\hline
\end{tabular}
\tablefoot{
\tablefoottext{a}{The numbers in brackets include possibly coincident cases ({\it P}).}
\tablefoottext{b}{including LS~5039}
}
\end{minipage}
\end{table}

\end{landscape}


\clearpage

\begin{thebibliography}{}

\bibitem[Abdo et al.(2007)]{milagro_GPS} Abdo, A.~A., Allen, B., Berley, D., et al. (MILAGRO Collaboration)\ 2007, \apjl, 664, L91
\bibitem[Abdo et al.(2009a)]{bsl_lat} Abdo, A. A., Ackermann, M., Ajello, M., et al. (Fermi/LAT Collaboration)\ 2009a, \apjs, 183, 46
\bibitem[Abdo et al.(2009b)]{latpsr_blind} Abdo, A. A., Ackermann, M., Ajello, M., et al. (Fermi/LAT Collaboration)\ 2009b, Science, 325, 840
\bibitem[Abdo et al.(2009c)]{lat_velapsr} Abdo, A. A., Ackermann, M., Atwood, W. B., et al. (Fermi/LAT Collaboration)\ 2009c, \apj, 696, 1084
\bibitem[Abdo et al.(2009d)]{lat_LSI61} Abdo, A. A., Ackermann, M., Ajello, M., et al. (Fermi/LAT Collaboration)\ 2009d, \apjl, 701, L123
\bibitem[Abdo et al.(2009e)]{lat_W51C} Abdo, A. A., Ackermann, M., Ajello, M., et al. (Fermi/LAT Collaboration)\ 2009e, \apjl, 706, L1
\bibitem[Abdo et al.(2009f)]{lat_ls5039} Abdo, A. A., Ackermann, M., Ajello, M., et al. (Fermi/LAT Collaboration)\ 2009f, \apjl, 706, L56
\bibitem[Abdo et al.(2009g)]{milagro_bsl} Abdo, A. A., Allen, B.~T., Aune, T., et al. (MILAGRO Collaboration)\ 2009g, \apjl, 700, L127
\bibitem[Abdo et al.(2010)]{lat_psr_cat} Abdo, A.~A., Ackermann, M., Ajello, M., et al. (Fermi/LAT Collaboration)\ 2010, \apjs, in press, preprint[arXiv:0910.1608]
\bibitem[Acciari et al.(2009a)]{veritas_LSI61} Acciari, V.~A., Aliu, E., Arlen, T., et al. (VERITAS Collaboration)\ 2009a, \apj, 700, 1034
\bibitem[Acciari et al.(2009b)]{veritas_ic443} Acciari, V.~A., Aliu, E., Arlen, T., et al. (VERITAS Collaboration)\ 2009b, \apjl, 698, L133
\bibitem[Acciari et al.(2009c)]{veritas_2227} Acciari, V.~A., Aliu, E., Arlen, T., et al. (VERITAS Collaboration)\ 2009c, \apjl, 703, L6
\bibitem[Acero et al.(2010)]{hess_GC_pos} Acero, F., Aharonian, F. A., Akhperjanian, A.~G., et al. (H.E.S.S. Collaboration)\ 2010, \mnras, 402, 1877
\bibitem[Aharonian et al.(2005a)]{aha05} Aharonian, F. A., Akhperjanian, A.~G., Aye, K.-M., et al. (H.E.S.S. Collaboration)\ 2005a, \aap, 436, L17
\bibitem[Aharonian et al.(2005b)]{hegra_TeV2032} Aharonian, F. A., Akhperjanian, A.~G., Beilicke, M., et al. (HEGRA Collaboration)\ 2005b, \aap, 431, 197
\bibitem[Aharonian et al.(2006a)]{aha06a} Aharonian, F. A., Akhperjanian, A.~G., Bazer-Bachi, A.~R., et al. (H.E.S.S. Collaboration)\ 2006a, \aap, 448, L19
\bibitem[Aharonian et al.(2006b)]{hess_velaX} Aharonian, F. A., Akhperjanian, A.~G., Bazer-Bachi, A.~R., et al. (H.E.S.S. Collaboration)\ 2006b, \aap, 448, L43
\bibitem[Aharonian et al.(2006c)]{hess_EBL_nature} Aharonian, F. A., Akhperjanian, A.~G., Bazer-Bachi, A.~R., et al. (H.E.S.S. Collaboration)\ 2006c, Nature, 440, 1018
\bibitem[Aharonian et al.(2006d)]{hess_crab} Aharonian, F. A., Akhperjanian, A.~G., Bazer-Bachi, A.~R., et al. (H.E.S.S. Collaboration)\ 2006d, \aap, 457, 899
\bibitem[Aharonian et al.(2006e)]{hess_LS5039} Aharonian, F. A., Akhperjanian, A.~G., Bazer-Bachi, A.~R., et al. (H.E.S.S. Collaboration)\ 2006e, \aap, 460, 743
\bibitem[Aharonian et al.(2006f)]{hess_survey} Aharonian, F. A., Akhperjanian, A.~G., Bazer-Bachi, A.~R., et al. (H.E.S.S. Collaboration)\ 2006f, \apj, 636, 777
\bibitem[Aharonian et al.(2006g)]{hess_kookaburra} Aharonian, F. A., Akhperjanian, A.~G., Bazer-Bachi, A.~R., et al. (H.E.S.S. Collaboration)\ 2006g, \aap, 456, 245
\bibitem[Aharonian et al.(2006h)]{hess_GC_DarkMatter} Aharonian, F. A., Akhperjanian, A.~G., Bazer-Bachi, A.~R., et al. (H.E.S.S. Collaboration)\ 2006h, \prl, 97, 221102
\bibitem[Aharonian et al.(2007a)]{hess_westerlund2_1023} Aharonian, F. A., Akhperjanian, A.~G., Bazer-Bachi, A.~R., et al. (H.E.S.S. Collaboration)\ 2007a, \aap, 467, 1075
\bibitem[Aharonian et al.(2007b)]{hess_psr_1718_1809} Aharonian, F. A., Akhperjanian, A.~G., Bazer-Bachi, A.~R., et al. (H.E.S.S. Collaboration)\ 2007b, \aap, 472, 489
\bibitem[Aharonian et al.(2008a)]{aha08} Aharonian, F. A., Akhperjanian, A.~G., Barres de Almeida, U., et al. (H.E.S.S. Collaboration)\ 2008a, \aap, 477, 481
\bibitem[Aharonian et al.(2008b)]{CTB_37A} Aharonian, F. A., Akhperjanian, A.~G., Barres de Almeida, U., et al. (H.E.S.S. Collaboration)\ 2008b, \aap, 490, 685
\bibitem[Aharonian et al.(2008c)]{hess_1801-233} Aharonian, F. A., Akhperjanian, A.~G., Barres de Almeida, U., et al. (H.E.S.S. Collaboration)\ 2008c, \aap, 481, 401
\bibitem[Aharonian et al.(2008d)]{hess_dark} Aharonian, F. A., Akhperjanian, A.~G., Barres de Almeida, U., et al. (H.E.S.S. Collaboration)\ 2008d, \aap, 477, 353
\bibitem[Aharonian et al.(2008e)]{VHE_Review_08} Aharonian, F., Buckley, J., Kifune, T., \& Sinnis, G.\ 2008e, Reports on Progress in Physics, 71, 096901
\bibitem[Aharonian et al.(2009)]{hess_1908+063} Aharonian, F. A., Akhperjanian, A.~G., Anton, G., et al. (H.E.S.S. Collaboration)\ 2009, \aap, 499, 723
\bibitem[Albert et al.(2007)]{magic_ic443} Albert, J., Aliu, E., Anderhub, H., et al. (MAGIC Collaboration)\ 2007, \apjl, 664, L87
\bibitem[Albert et al.(2008)]{magic_crab_nebula} Albert, J., Aliu, E., Anderhub, H., et al. (MAGIC Collaboration)\ 2008, \apj, 674, 1037
\bibitem[Albert et al.(2009)]{magic_LSI61} Albert, J., Aliu, E., Anderhub, H., et al. (MAGIC Collaboration)\ 2009, \apj, 693, 303
\bibitem[Aliu et al.(2008)]{magic_crab_psr} Aliu, E., Anderhub, H., Antonelli, L. A., et al. (MAGIC Collaboration)\ 2008, Science, 322, 1221
\bibitem[Aliu et al.(2009)]{veritas_psr_limit_HDGS} Aliu, E., for the VERITAS Collaboration\ 2009, AIP Conference Proceedings, 1085, 324
\bibitem[Atwood et al.(2009)]{lat_technical} Atwood, W.~B., Abdo, A.~A., Ackermann, M., et al. (Fermi/LAT Collaboration)\ 2009, \apj, 697, 1071
\bibitem[Berge et al.(2007)]{berge07} Berge, D., Funk, S., \& Hinton, J. A. 2007, \aap, 466, 1219
\bibitem[Bertsch et al.(1992)]{Geminga_psr_discovery} Bertsch, D.~L., Brazier, K.~T.~S., Fichtel, C.~E., et al.\ 1992, \nat, 357, 306
\bibitem[Bertsch et al.(2000)]{Bertsch00} Bertsch, D.~L., Hartman, R. C., Hunter, S. D., et al.\ 2000, American Institute of Physics Conference Series, 510, 504
\bibitem[Celik et al.(2009)]{lat_geminga} Celik, O., on behalf of the Fermi-LAT Collaboration\ 2009, to be published in the proceedings of the 31st International Cosmic Ray Conference
\bibitem[Chaves et al.(2009a)]{hess_survey09} Chaves, R.~C.~G., on behalf of the H.E.S.S. Collaboration\ 2009a, to be published in the proceedings of the 31st International Cosmic Ray Conference
\bibitem[Chaves et al.(2009b)]{hess_1848_HDGS} Chaves, R.~C.~G., de O{\~n}a Wilhemi, E., \& Hoppe, S., for the H.E.S.S. Collaboration\ 2009b, AIP Conference Proceedings, 1085, 372
\bibitem[Djannati-Atai et al.(2008a)]{icrc07_1908+063} Djannati-Atai, A., de Jager, O.~C., Terrier, R., Gallant, Y.~A., Hoppe, S., for the H.E.S.S. Collaboration,\ 2008a, Proceedings of the 30th International Cosmic Ray Conference; Universidad Nacional Aut\'onoma de M\'exico, Mexico City, Mexico, 2, 823
\bibitem[Djannati-Atai et al.(2008b)]{icrc07_1833_105} Djannati-Atai, A., O\~na-Wilhelmi, E., Renaud, M., Hoppe, S., for the H.E.S.S. Collaboration,\ 2008b, Proceedings of the 30th International Cosmic Ray Conference; Universidad Nacional Aut\'onoma de M\'exico, Mexico City, Mexico, 2, 863
\bibitem[Dubois et al.(2009)]{hess_velaX_2009} Dubois, F., Gl\"uck, B., de Jager, O. C., et al. 2009, Proceedings of the 31st ICRC, {\L}\'od\'z, Poland
\bibitem[Feldman \& Cousins(1998)]{feldman98} Feldman, G. J., \& Cousins, R. D. 1998, Phys. Rev. D., 57, 3873
\bibitem[Fiasson et al.(2009)]{hess_w51} Fiasson, A., Marandon, V., Chaves, R.~C.~G., Tibolla, O., for the H.E.S.S. Collaboration\ 2009, to be published in the proceedings of the 31st International Cosmic Ray Conference
\bibitem[Finnegan et al.(2009)]{veritas_geminga} Finnegan, G., on behalf of the VERITAS Collaboration\ 2009, to be published in the proceedings of the 31st International Cosmic Ray Conference
\bibitem[Funk et al.(2004)]{funk04} Funk, S., Hermann, G., Hinton, J. A. et al.\ 2004, Astropart. Phys., 22, 285
\bibitem[Funk et al.(2008)]{funk08} Funk, S., Reimer, O., Torres, D. F., \& Hinton, J. A.\ 2008, \apj, 679, 1299
\bibitem[Gargano et al.(2009)]{lat_egr_psr} Gargano, F. for the Fermi/LAT Collaboration\ 2009, to be published in the proceedings of the 31st International Cosmic Ray Conference
\bibitem[Grondin et al.(2009)]{lat_crab} Grondin, M.-H., on behalf of the Fermi-LAT Collaboration\ 2009, to be published in the proceedings of the 31st International Cosmic Ray Conference
\bibitem[Halpern et al.(2001)]{PSR2229_discovery_halpern01} Halpern, J.~P., Camilo, F., Gotthelf, E.~V., et al.\ 2001, \apjl, 552, L125
\bibitem[Hartman et al.(1999)]{3rd_egret_cat} Hartman, R.~C., Bertsch, D.~L., Bloom, S.~D., et al. 1999, \apjs, 123, 79
\bibitem[Hillas(1996)]{hillas96} Hillas, A. M. 1996, \ssr, 75, 17
\bibitem[Hoppe et al.(2008)]{hess_survey_2007} Hoppe, S., for the H.E.S.S. Collaboration,\ 2008, Proceedings of the 30th International Cosmic Ray Conference; Universidad Nacional Aut\'onoma de M\'exico, Mexico City, Mexico, 2, 579, preprint[arXiv:0710.3528]
\bibitem[Hoppe et al.(2009)]{hess_1706} Hoppe, S., de O{\~n}a Wilhemi, E., Kh{\'e}lifi, B., Chaves, R.~C.~G., de Jager, O.~C., Stegmann, C., Terrier, R., for the H.E.S.S.~Collaboration\ 2009, to be published in the proceedings of the 31st International Cosmic Ray Conference, preprint[arXiv:0906.5574]
\bibitem[Hinton(2009)]{Hinton09_review} Hinton, J.\ 2009, New Journal of Physics, 11, 055005
\bibitem[de Jager et al.(1996)]{egret_crab_nebula} de Jager, O.~C., Harding, A.~K., Michelson, P.~F., Nel, H.~I., Nolan, P.~L., Sreekumar, P., \& Thompson, D.~J.\ 1996, \apj, 457, 253
\bibitem[Lamb \& Macomb(1997)]{GeV_cat} Lamb, R.~C., \& Macomb, D.~J.\ 1997, \apj, 488, 872
\bibitem[Lemoine-Goumard  et al.(2009)]{lat_velaX} Lemoine-Goumard, M., and Grondin, M.-H., on behalf of the Fermi-LAT Collaboration and the Pulsar Timing Consortium, 2009, to be published in the proceedings of the 31st International Cosmic Ray Conference
\bibitem[Li \& Ma(1983)]{LiMa83} Li, T.-P., \& Ma, Y.-Q.\ 1983, \apj, 272, 317
\bibitem[Mayer-Hasselwander et al.(1994)]{egret_geminga} Mayer-Hasselwander, H.~A., Bertsch, D.~L., Brazier, K.~T.~S., et al.\ 1994, \apj, 421, 276
\bibitem[Ng et al.(2005)]{ng_chandra_rabbit} Ng, C.-Y., Roberts, M.~S.~E., \& Romani, R.~W.\ 2005, \apj, 627, 904
\bibitem[Ohm et al.(2009)]{hess_westerlund1_1648} Ohm, S., Horns, D., Reimer, O., et al., for the H.E.S.S. Collaboration\ 2009, HEPIMS Workshop (Jaen 2009), to be published in the Publications of the Astronomical Society of the Pacific, preprint[arXiv:0906.2637]
\bibitem[Ong et al.(2009)]{VERITAS_results_Fermi_Sym09} Ong, R.~A., Acciari, V. A., Arlen, T., for the VERITAS collaboration\ 2009, to be published in the proceedings of the 2009 Fermi Symposium, preprint[arXiv:0912.5355]
\bibitem[Reimer et al.(2008)]{reimar_icrc2007} Reimer, O., Funk, S., Torres, D.~F., \& Hinton, J.\ 2008, Proceedings of the 30th International Cosmic Ray Conference; Universidad Nacional Aut\'onoma de M\'exico, Mexico City, Mexico, 2, 613
\bibitem[de los Reyes et al.(2009)]{magic_psr_limits} de los Reyes, R., Bednarek, W., Camara, M., \& Lopez, M., for the MAGIC Collaboration\ 2009, to be published in the proceedings of the 31st International Cosmic Ray Conference
\bibitem[Thompson et al.(1996)]{egret_psr1706} Thompson, D.~J., Bailes, M., Bertsch, D.~L., et al.\ 1996, \apj, 465, 385
\bibitem[Tibolla et al.(2009a)]{HowCanFermiHelp_SciNeGHe08} Tibolla, O., on behalf of the H.E.S.S. Collaboration\ 2009a, AIP Conference Proceedings, 1112, 211
\bibitem[Tibolla et al.(2009b)]{hess_1741_HDGS} Tibolla, O., Komin, N., Kosack, K., \& Naumann-Godo, M., on behalf of the H.E.S.S. Collaboration\ 2009b, AIP Conference Proceedings, 1085, 249
\bibitem[Weinstein et al.(2009)]{VERITAS_cygnus_survey09} Weinstein, A., for the VERITAS collaboration\ 2009, to be published in the proceedings of the 2009 Fermi Symposium, preprint[arXiv:0912.5355]
\bibitem[Zanin et al.(2009)]{MAGIC_results_Fermi_Sym09} Zanin, R., for the MAGIC collaboration\ 2009, to be published in the proceedings of the 2009 Fermi Symposium, preprint[arXiv:0912.3671]
\bibitem[Zepka et al.(1996)]{PSR0631_discovery} Zepka, A., Cordes, J.~M., Wasserman, I., \& Lundgren, S.~C.\ 1996, \apj, 456, 305

\end{thebibliography}
\end{document}